\documentclass{aa}
\usepackage{graphicx}
\usepackage{txfonts}
\usepackage{natbib}
\usepackage{url}

\newcommand{\arsec}{'\!\hskip0.4pt'\hspace{-0.2mm}}

\newcommand{\dotarsec}{.\hspace{-0.9mm}'\!\hskip0.4pt'\hspace{-0.2mm}}
\newcommand{\ve}[1]{{\rm \bf {#1}}}

\begin{document}
   \title{The continuum intensity as a function of magnetic field}
   \subtitle{I. Active region and quiet Sun magnetic elements}

   \author{P. Kobel \inst{1} \and S. K. Solanki \inst{1,2} \and J. M. Borrero \inst{3} }

   \offprints{P. Kobel}

   \institute{Max-Planck Institut f\"{u}r Sonnensystemforschung, Max-Planck-Stra\ss e 2, 37191 Katlenburg-Lindau, Germany\\
              \email{philippe.kobel@a3.epfl.ch}
              \and School of Space Research, Kyung Hee University, Yongin, Gyeonggi, 446-701, Korea \\
              \and Kiepenheuer-Institut f\"{u}r Sonnenphysik, Sch\"{o}neckstr.6, 79104 Freiburg, Germany }

   \date{...}


  \abstract
      {Small-scale magnetic fields are major contributors to the solar irradiance variations. Hence, the
      continuum intensity contrast of magnetic elements in the quiet Sun (QS) network and in active region (AR)
      plage is an essential quantity that needs to be measured reliably.}
      {By using Hinode/SP disk center data at a constant, high spatial resolution, we aim at updating results of
      earlier ground-based studies of contrast vs. magnetogram signal, and to look for systematic differences
      between AR plages and QS network.}
      {The field strength, filling factor and inclination of the field was retrieved by means of a
Milne-Eddington inversion (VFISV code). As in earlier studies, we then performed a pixel-by-pixel study of
630.2 nm continuum contrast vs. apparent (i.e. averaged over a pixel) longitudinal magnetic field over large fields of
view in ARs and in the QS.}
      {The continuum contrast of magnetic elements reaches larger values in the QS (on average 3.7\%)
than in ARs (on average 1.3\%). This could not be attributed to any systematic difference in the
chosen contrast references, so that it mainly reflects an intrinsic brightness difference.
At Hinode's spatial resolution, the relationship between contrast and
apparent longitudinal field strength exhibits a peak at around 700 G in both the QS and ARs, whereas earlier
lower resolution studies only found a peak in the QS and a monotonous decrease in ARs. We attribute this
discrepancy both to our careful removal of the pores and their close surroundings affected by the telescope
diffraction, as well as to the enhanced spatial resolution and very low scattered light of the Hinode Solar
Optical Telescope. We verified that the magnetic elements producing the peak in the contrast curve are rather
vertical in the AR and in the QS, so that the larger contrasts in the QS cannot be explained by larger
inclinations, as had been proposed earlier.
}
 {According to our inversions, the magnetic elements producing the peak of the contrast curves have similar
properties (field strength, inclination, filling factor) in ARs and in the QS, so that the larger
brightness of magnetic elements in the QS remains unexplained. Indirect evidence suggests that the
contrast difference is not primarily due to any difference in average size of the magnetic elements. A possible
explanation lies in the different efficiencies of convective energy transport in the QS and in ARs, which will be
the topic of a second paper.}


   \titlerunning{The continuum intensity as a function of magnetic field I.}
   \authorrunning{P. Kobel et al.}

   \maketitle


\section{Introduction}
\label{sec_intro}

The quiet Sun (QS) network and active region (AR) plages are the two most prominent components of solar
photospheric magnetism outside Sunspots \citep{Solanki_rev06}. Both components contain small-scale magnetic
features possessing kG field strengths, which appear bright in spectral line cores and theoretically also at
continuum wavelengths, even at disk center \citep[][]{Deinzer84, Schuessler88, Knoelker88, Knoelker91,
Voegler03}. Their excess brightness is due to the fact that they are hotter than their surroundings at equal
optical depth \citep[see][]{Schuessler_rev92}. Owing to their enhanced brightness, these so-called ``magnetic
elements'' are key players in the total solar irradiance variations on the timescale of days to the solar cycle
\citep{Krivova03, Domingo09} and very likely also longer \citep{Krivova07}. Yet the observed continuum brightness
of magnetic elements in ARs and in the QS is still a matter of debate, as is the relative contribution of ARs and
QS network to the total solar irradiance variations \citep[although the variation in area and Ca II K emission of
ARs were found to dominate on timescales of a solar cycle,][]{Walton03}. It is thus fundamental to quantify the
continuum brightness of magnetic elements in the QS network and in ARs separately, inasmuch as visible continuum
wavelengths contribute $\sim 50$ \% to the total solar irradiance and $\sim 30 \%$ to its solar cycle variation
\citep[according to the calculations of][]{Krivova06}.

The brightness of magnetic features is usually not measured in absolute sense, but relative to the mean intensity of a
reference quiet photosphere, i.e. by their ``contrast''.
The contrast at continuum wavelengths is thus directly related to the temperature excess with
respect to this quiet photosphere at the level $\tau = 1$.
At disk center (and generally at any fixed heliocentric angle), the temperature excess of a magnetic
feature depends on its field strength, which determines the depth of the opacity depression, and on the
effectiveness of the radiative heating from its ``hot walls'' \citep{Spruit76}. This latter depends on the size
of the flux tube, in particular the ratio of the surface of the walls to the internal volume, as well as on the
efficiency of the surrounding convective heat transport.
Since measurements based on line-ratio techniques and inversions indicate similar kG field strengths for magnetic
elements in network and plages \citep[with a weak dependence on the filling factor,][]{Frazier72, Stenflo73,
Solanki84, Stenflo85, Zayer90, Rabin92a, Rabin92b}, the continuum contrast of magnetic elements at disk center should be
primarily dictated by their sizes \citep{Spruit81}. Note that the much weaker equipartition fields
detected mostly in the internetwork by, e.g. \citet{Lin95, Solanki96, Lites02, Khomenko03} and the ubiquitous
horizontal fields detected with Hinode \citep{Lites08, Ishikawa09} are likely to display a much smaller, possibly
unimportant contrast \citep{Schnerr10}. A recent study by \citet{Viticchie10} also suggests that the contrast of
disk center G-band bright points, which are associated with magnetic elements, mainly depends on their size while the
$\sim$kG field strength is rather constant.
Note that the contrast of magnetic elements also increases with the heliocentric angle \citep[with an
eventual maximum, see][for reviews]{Solanki_rev93, Steiner07}, as progressively more of the hot granular wall
limbward of the flux tubes becomes visible \citep[``hot wall effect'', see e.g.][]{Spruit76, Keller04} and the
optical depth shifts upward where magnetic elements have a larger temperature excess due to the shallower
temperature gradient inside than outside the flux tubes \citep{Steiner05}. Signatures of these hot granular walls
(``faculae'') can be seen directly in the high resolution continuum and G-band images of \citet{Lites04},
\citet{Kobel09} \citet{Hirz05} and \citet{Berger07}, with results of contrast at various heliocentric angles
being presented in the last two papers. While the present paper only deals with disk center data, the
center-to-limb variation of the continuum contrast of magnetic elements will be presented in a forthcoming one
\citep[][Paper III of this series]{Kobel11c}.
Finally, as observed by \citet{Berger96, Berger01} and \citet{DePontieu06}, the contrast of individual magnetic
elements is also indirectly a function of time, since both the physical parameters influencing the contrast
(magnetic element size, field strength and inclination) and the seeing can vary with time. Thus, the above
considerations hold statistically when averaging in time or over an ensemble of magnetic elements. Under these
conditions the dependence on the inclination of magnetic elements can be omitted because they are on average
close to vertical due to magnetic buoyancy.
To indirectly gain information about how the sizes of magnetic elements influence their contrasts at disk
center, one can statistically investigate the relation between contrast and ``magnetogram signal''
\citep[i.e. net longitudinal field in the resolution element obtained from the calibration of Stokes $V$, see
][]{Stenflo_rev08}. Since the kG flux tubes in QS and in AR are rather vertical and have similar field strengths,
at a fixed disk position the magnetogram signal should only scale with the fractional area of the
resolution element filled by magnetic fields, i.e. the ``filling factor''. At disk center, this filling
factor reflects the total cross section (at the height of line formation) of the magnetic features present in the
resolution element.

The easiest and most straightforward way to carry out such an analysis is to make scatterplots of the
contrast vs. magnetogram signal \citep[cf.][]{Frazier71}, and average the contrasts in bins of magnetogram
signal. Using this method, \citet{Title92} and \citet{Topka92} found that the average continuum contrast
(at 676.8, 525, 557.6 and 630.2 nm) of magnetic features in ARs at disk center was negative for all bins
of magnetogram signal, i.e. their continuum brightness was never greater than the mean quiet photosphere. To be
consistent with the flux tube models predicting intrinsically bright magnetic elements, the authors invoked the
effect of limited spatial resolution, smearing the magnetic elements with surrounding dark moats \citep{Title92,
Topka92} and intergranular lanes \citep{Title96}. Applying the same method, but on QS network data (of similar
resolution), \citet{Lawrence93} found that the average contrast reaches positive values at magnetogram
signals at which the AR contrast was negative. These authors then proposed that the larger contrasts in the QS
could be explained by more inclined flux tubes, whereby their bright hot walls would be better visible
\citep{Lawrence93}, but left this hypothesis unverified.

A drawback of the pixel-by-pixel analysis used in the aforementioned works is that it does not preserve
the information about individual magnetic features. Instead, the contrast of pixels coming from unrelated
features (especially if these features are composed of several pixels) can be assigned to the same bin of
magnetogram signal, with the risk of blending information from bright magnetic elements with signals of
micropores and intergranular lanes. To focus solely on magnetic elements, an alternative method is to segment the
bright features in images and compare the average contrast and magnetogram value for each feature separately
\citep[as performed by e.g.][]{Viticchie10, Berger07}. However, the segmentation requires the use of joint
high-resolution filtergrams in molecular bands in which the feature contrast is enhanced like the G-band or CN
band, first used by \citet{Sheeley69} and \citet{Muller84} \citep[see also][for comparative studies of both bands
based on observations and simulations, respectively]{Zakharov07, Uitenbroek06}. Moreover, the segmentation
necessarily relies on some arbitrary thresholds, unlike the pixel-by-pixel method. Note that the two methods
yield distinct results also because the population of magnetic features they study are different: since the
segmentation-based method only selects bright magnetic features, the resulting contrast vs. magnetogram signal
curves tend to increase monotonically or reach saturation \citep{Viticchie10, Berger07}. In contrast, due to the
inclusion of all pixels, the contrast curves obtained by the second method rather exhibit a peak followed by a
decrease due to larger and darker magnetic features (e.g. Fig. \ref{fig_C_vs_B_FOV} herein).

In this paper we compare the 630.2 nm continuum contrast of magnetic elements in ARs and in the QS, using
data from the spectropolarimeter onboard Hinode, which allow good determinations of the magnetic field vector
and have higher spatial resolution (with constant image quality) than the data of Topka and colleagues (see
above). For a direct comparison with their studies we use the same pixel-by-pixel analysis method. We first
repeat scatterplots of continuum contrast vs. longitudinal magnetic field in ARs and in the QS, as presented
Sect. \ref{sec_CvsB}. In Section \ref{sec_incl} we compare the inclinations of magnetic elements in the AR and
the QS to test the hypothesis of \citet{Lawrence93}, while in Sect. \ref{sec_polarity} we also look for contrast
and inclination differences between the opposite polarities in ARs. We then check in Sect. \ref{sec_brightness}
that the results are not biased by a systematic difference between the contrast references in ARs and in the QS.
Finally, in Sect. \ref{sec_comp} we explain the qualitative discrepancies between our scatterplots in ARs and
those of previous studies. Our conclusions are presented and discussed in Sect. \ref{sec_concl}.


\section{Dataset analysis}

\subsection{Hinode/SP scans}
\label{sec_scans}

We selected an ensemble of 6 spectropolarimetric scans over active regions and 4 scans over the quiet Sun
performed very close to disk center (see Table \ref{table_scans}) by the Hinode/SP instrument \citep{HinodeSP,
HinodeSP2, Hinode07, HinodeSOT, HinodeSOT2}.

The SP delivers profiles of the four Stokes parameters (for every pixel along its slit) in a visible wavelength
range covering both the Fe I 630.15 nm and 630.25 nm lines, at a constant spatial resolution of $0\dotarsec3$
\citep[see e.g.][for more details]{Lites08}.

The selected scans were performed in the ``normal mode'', i.e. with an exposure time of 4.8 s resulting in
typical rms noise levels at disk center of $1.1 \times 10^{-3}$ and $1.2 \times 10^{-3}$ for Stokes $V$ and
$Q,U$, respectively (in units of continuum intensity $I_{\rm c}$). All the profiles were calibrated via the
\verb"sp_prep" routine of the SolarSoft package
\footnote{\url{http://www.lmsal.com/solarsoft/sswdoc/index_menu.html}}. Note that although the point spread
function of the SOT/SP has been calculated by \citet{Danilovic08}, no inversions of this function are currently
available so that the calibrated data were not further processed.

\begin{table}
\caption{List of the SP scans used in this work. $t_{\rm start}$ denotes the starting time of the scans.}
\vskip3mm \centering
\begin{tabular}{c c c c}     
\hline\hline
date (dd-mm-yy) & $t_{\rm start}$ (UT) & target & NOAA  \\
\hline
11-12-06 & 13:10:09 & AR & 10930\\
05-01-07 & 11:20:09 & AR & 10933\\
01-02-07 & 12:14:05 & AR & 10940\\
28-02-07 & 11:54:34 & AR & 10944\\
01-05-07 & 21:00:06 & AR & 10953\\
11-05-07 & 12:35:53 & AR & 10955\\
\hline
10-03-07 & 11:37:36 & QS & --\\
23-04-07 & 11:14:06 & QS & --\\
24-04-07 & 01:21:04 & QS & --\\
27-04-07 & 08:50:03 & QS & --\\
\hline
\end{tabular}
\label{table_scans}
\end{table}

\subsection{Maps of continuum intensity and heliocentric distance}
\label{sec_Ic_mu}

Maps of the continuum intensity $I_c$ (calculated in the red continuum of the 630.2 nm line) were provided by the
\verb"sp_prep" procedure. Because of the short exposure times compared to the granulation lifetime ($\sim 10$ min),
these continuum maps can be thought of as quasi-instantaneous images over small distances \citep[cf.][]{Lites08}.
They are thus appropriate for the study of small-scale magnetic features, whose evolution timescales are roughly
comparable with those of the granulation \citep{Berger96,Berger98}.

\verb"sp_prep" also calculates the right ascension $x$ and declination $y$ (heliocentric cartesian coordinates),
and thereby the $\mu = cos \theta$ value at each pixel of the maps (where $\theta$ is the heliocentric angle). In
the present study, these ``$\mu$ maps'' were used to select portions of the scans located at the very disk
center, i.e. where $\mu >0.99$. Only these portions are studied here.

No correction of $I_c$ for limb darkening was performed, as the latter can be estimated to only $\sim0.05\%$
between $\mu=1$ and $\mu=0.99$ (using the $5^{\rm th}$-order polynomial in $\mu$ published by
\citet{Neckel94}), and therefore is considered negligible.

\subsection{Inversions}
\label{sec_inversions}


The observed Stokes spectra at each spatial pixel were inverted with the VFISV (Very Fast Inversion of the Stokes Vector)
code of \citet{Borrero10}. We refer to this article for all details. This code produces synthetic Stokes profiles of the
630.2 nm line using the Milne-Eddington solution (M-E) for the radiative transfer equation \citep[see, e.g.,][]{Deltoro03},
as follows:

\begin{equation}
\ve{I}^{syn}(\lambda) = \alpha \ve{I}_{\rm mag}(\lambda,\mathcal{X}) + (1-\alpha) \ve{I}_{\rm nmag}(\lambda) \;,
\end{equation}

\noindent where $\ve{I}_{\rm mag}(\lambda,\mathcal{X})$ refers to the Stokes vector arising from the magnetized
part of the pixel. Thus, the set of parameters $\mathcal{X}$ refers to the physical properties of the magnetized
plasma which, under the M-E approximation, are:

\begin{equation}
\mathcal{X}= [S_0, S_1, \eta_0, a, \Delta\lambda_D, B, \gamma, \phi, v_{\rm los}, \alpha] \;,
\end{equation}

\noindent where the first 5 are the thermodynamic parameters: source function, $S_0$, source function gradient,
$S_1$, ratio of the absorption coefficient between the continuum and line-center, $\eta_0$, damping parameter
$a$, and Doppler width, $\Delta\lambda_D$.\footnote{Note that $a$ was maintained at a prescribed value.} The
next three refer to the three components of the magnetic field vector: intrinsic field strength, $B$, inclination
of the magnetic field vector with respect to the observer $\gamma$, and azimuthal angle in the plane
perpendicular to the observer's, $\phi$. $v_{\rm los}$ denotes the line-of-sight (LOS) velocity of the magnetized
plasma. Finally, the ``filling factor'' $\alpha$ is a geometrical parameter that refers to the fractional amount
of light that corresponds to the magnetized plasma.

To retrieve $\alpha$, we adopted the ``local stray light'' approach used by \citet{Orozco07} \citep[see
also][]{BorreroKobel11}. $\ve{I}_{\rm nmag}(\lambda)$ refers to the Stokes vector arising from the non-magnetized
portion of the pixel, and therefore we consider it not to be polarized: $\ve{I}_{\rm nmag}(\lambda)=(I_{\rm
nmag},0,0,0)$. $I_{\rm nmag}$ is obtained by averaging the intensity profiles from the neighboring pixels within
$1\arsec$ of the inverted one. Note that, because of the way it is constructed, $I_{\rm nmag}$ cannot distinguish
between non-magnetized plasma within the resolution element, the effects of the telescope diffraction and the
scattered light within the instrument. However, given the small amount of scattered light in Hinode's
spectropolarimeter \citep{Danilovic08} the last contribution can be neglected.

We have observationally determined that the noise for Stokes $I$ is typically between a factor of 5 and 10 larger
than in $Q,U$ or $V$ due to the flatfield correction (Lites, private communication), which is only known to the
order of 1 \%. To account for that, Stoke $I$ was given 5 times lower weight than the polarization profiles in
the inversion scheme.

All the pixels were inverted (no polarization selection) and included in the analysis, as the latter mainly deals
with the longitudinal component of the field, which can be considered reliable over the range corresponding to the
magnetic elements having Stokes $V$ signal largely above the noise (see Sect. \ref{sec_CvsB}).

In contrast to older studies using magnetograms (see Sec. \ref{sec_intro}), the present approach has the
advantage of remaining valid in all regimes of intrinsic field strength, as it is based on analytically solving
the full radiative transfer equation. Magnetograms rely on a calibration of Stokes $V$ \citep[cf.][]{Topka92}
that only holds in the ``weak field regime'', and thus the values of the magnetogram signal are underestimated
for vertical fields approaching 1000 G in intrinsic strength \citep{Innocenti04}, which is the so-called ``Zeeman
saturation effect'' \citep{Howard72, Stenflo73}.


\section{Results}

\subsection{Scatterplots of contrast vs. longitudinal field for active region and quiet Sun}
\label{sec_CvsB}

\begin{figure*}[hbt]
\centering
\includegraphics[width=0.8\textwidth]{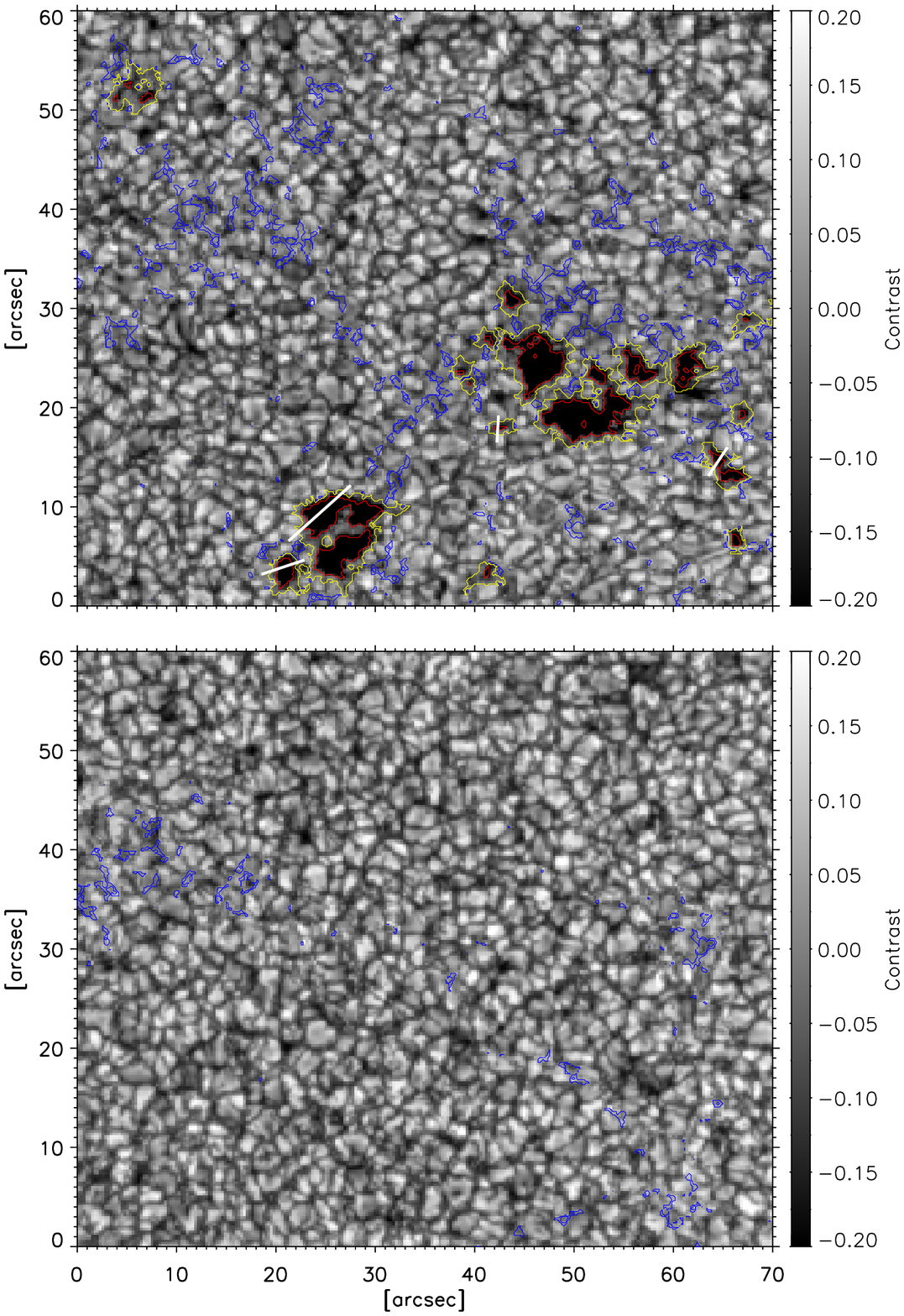}
\caption{Upper panel: Continuum contrast of an
active region plage field of view at disk center ($\mu > 0.99$), extracted from the SP scan of day 01-02-2007 (see Table
\ref{table_scans}).
Bottom panel: Same for a field of view including quiet Sun network, extracted from the SP scan of day 24-04-2007.
Blue contours: locations of the pixels where $B_{\rm app,los}$ lies in an interval of $\pm 200$ G around the peak
value of the contrast vs. $B_{\rm app,los}$ relation (see Fig. \ref{fig_C_vs_B_FOV}). The red contours surround
the ``core'' of the pores where the contrast is below $-0.15$ and $B_{\rm app,los} > 900$ G, while the yellow
contours outline the entire pore areas removed from the analysis. The white lines across some pores coincide with
the locations of the cuts discussed in Sect. \ref{sec_comp}.}
\label{fig_FOV}
\end{figure*}

To start with, we performed a pixel-by-pixel comparison of continuum contrast vs. longitudinal field \citep[as
undertaken by][hereafter TTL]{Title89, Topka92, Lawrence93} over active regions (ARs) and the quiet Sun (QS), to
see if and how the results would differ between these two targets at the constantly high spatial resolution of
Hinode.

Like TTL, we considered rather large fields of view (FOVs) of $70\arsec \times 60\arsec$. We analyzed a total of
10 FOVs containing QS network and 9 FOVs of AR plages, all at $\mu > 0.99$, extracted from the different SP scans
(see Sect. \ref{sec_scans}). As we obtained similar results with all FOVs (summarized in Table
\ref{table_fovs}), we restrict ourselves here to presenting the results for one such FOV centered on an AR plage
(see Fig. \ref{fig_FOV} upper panel), and one including QS network (Fig. \ref{fig_FOV} lower panel).

To simulate the magnetogram signal used by TTL, we considered the ``apparent'' strength of the longitudinal
component of the field $B_{\rm app,los} = B \alpha |$cos$\gamma|$ (see Sect. \ref{sec_inversions} for the
description of the variables $B$, $\alpha$ and $\gamma$). Like the magnetogram signal, it can be considered to
scale with the size of unresolved magnetic elements (see Sect. \ref{sec_intro}). We use the terminology
``apparent'' field strength to distinguish from the intrinsic one, since $B_{\rm app,los}$ is averaged over the
pixel. The continuum contrast at each pixel location (x,y) of a given FOV was defined as:
\begin{equation}
\label{eq_contrast}
{\rm Contrast(x,y)} = \frac{I_c(x,y) - \left< I_c \right>_{\rm ref, FOV}}{\left< I_c \right>_{\rm ref, FOV},}
\end{equation}
where $\left< I_c \right>_{\rm ref, FOV}$ is the contrast reference, taken as the mean continuum intensity of the
pixels having $B_{\rm app,los} < 100$ G in the FOV (corresponding to rather normal granulation).

\begin{figure}
\centering
\includegraphics[width=\columnwidth]{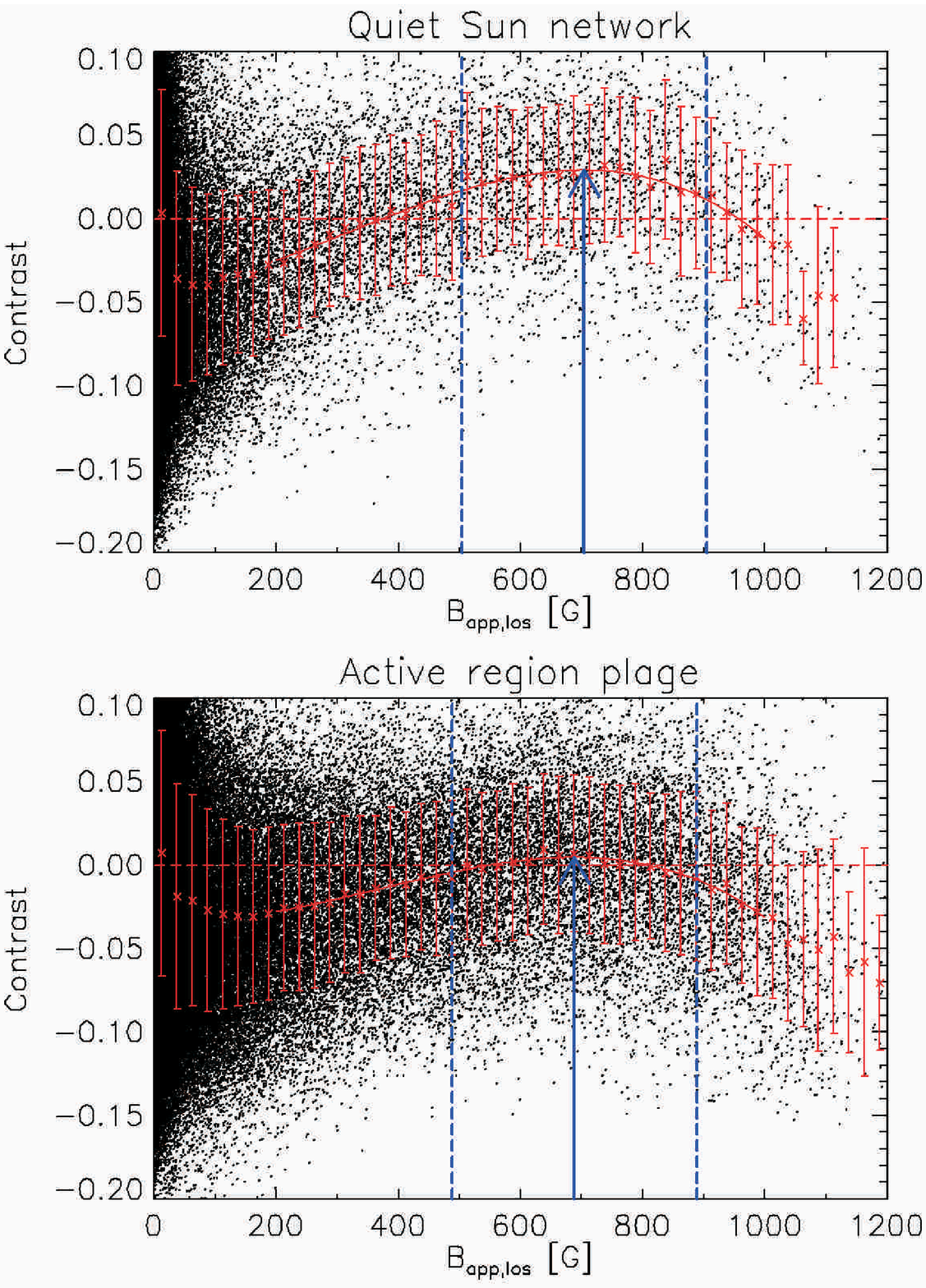}
\caption{Scatterplot of the continuum contrast vs. apparent longitudinal field strength $B_{\rm app,los}$ for the
quiet Sun (upper panel) and plage (bottom panel) fields of view shown in Figs. \ref{fig_FOV}, pores excluded. The
contrast reference (indicated by the horizontal dashed red line) is the mean intensity of the pixels where
$B_{\rm app,los} < 100$ G. Red crosses: average values of the continuum contrast inside $B_{\rm app,los}$-bins of
25 G width. The red error bars are the standard deviations inside each bin. Solid red curves are third-order polynomial
fits of the average values in the range 200 G $< B_{\rm app,los} < 1000$ G. The blue arrows indicate the maxima
of the fits, locating the ``peak'' of the contrast trends. The vertical dashed lines delimit the interval $\pm200$ G around the peak.}
\label{fig_C_vs_B_FOV}
\end{figure}

Figure \ref{fig_C_vs_B_FOV} displays the resulting scatterplots of the continuum contrast vs. $B_{\rm app,los}$
for the AR and for the QS FOVs shown in Fig. \ref{fig_FOV}.
To prevent pores from contaminating the contrasts in this range, we removed them according to the following
procedure. First, their inner dark cores were detected as any group of at least 4 pixels (the minimum number
corresponding to spatial resolution) having contrast below $-0.15$ and $B_{\rm app,los}>900$ G (assuming pores to
be resolved and thus with fields close to kG), such as enclosed by the red contours in Fig.
\ref{fig_FOV}. Next, the surroundings of the pores suffering from spurious magnetic signal due to the telescope
diffraction (see Sect. \ref{sec_comp}) were also eliminated by spatially extending the detected cores until
$B_{\rm app,los}$ drops below 200 G, which corresponds to the larger yellow contours in Fig. \ref{fig_FOV}.
This is a conservative procedure, which might assign somewhat too many pixels to pores, but importantly for our
work, it ensures that no pixels influenced by a pore's magnetic field are accidently assigned to magnetic elements.

To detect a trend within the scatterplots, the pixel contrasts were averaged into bins of $B_{\rm app,los}$ (with
a binwidth of 25 G), and a third-order polynomial was fitted to the average values of $B_{\rm app,los}$ between
200 G and $1000$ G.\footnote{We consider that $B_{\rm app,los}$ is reliably retrieved in the range $B_{\rm
app,los} > 200$ G, as more than 99.9 \% of the pixels in that range have a Stokes $V$ amplitude above 4.5 times
the rms noise level.} In these scatterplots, one can notice the large contrast scatter for $B_{\rm app,los} \sim
0$ due to granulation, while the scatter and the average contrasts decrease at the same time that the field
concentrates in intergranular lanes \citep[see][for more details on this part]{Schnerr10}. For the range $B_{\rm
app,los} > 200$ G, dominated by flux concentrations, the average contrast increases in both plots until reaching
a peak for some value of $B_{\rm app,los}$, corresponding to the maximum of the fits. For values of $B_{\rm
app,los}$ below the peak value, the features are on average less bright than the peak contrast, either because
their field strength is too low or because they are partially unresolved. For $B_{\rm app,los}$ larger than the
peak value, the features become progressively darker as the filling factor increases. Note that unlike our
trends, all the trends of TTL in ARs are monotonically decreasing. As explained in the Sect. \ref{sec_comp}, this
is not only due to our higher spatial resolution but mainly to our complete removal of pores.

Two qualitative observations can be made. Firstly, both the trends of the contrast vs. $B_{\rm app,los}$ in the
QS and in AR peak at a similar value $B_{\rm app,los} \sim 700$ G, as indicated by the arrows in Fig.
\ref{fig_C_vs_B_FOV} (and listed in Table \ref{table_fovs} for all the FOVs analyzed), whereas the AR trends of
TTL were monotonically decreasing. Secondly, even at Hinode's constant and high spatial resolution, the QS network
reaches larger continuum contrasts than in AR plage, in agreement with the findings of TTL. The average
of the peak contrasts in the QS FOVs is 3.7\% and 1.3\% in the AR FOVs, corresponding to a relative
difference of 2.4\%. Note that due to our higher spatial resolution (about twice higher than that of TTL, see
Sect. \ref{sec_comp}), the average contrasts reported here are larger than the ones measured by TTL: our QS peak
contrast reaches almost 3\% (and up to 6.2\% in other FOVs, see Table \ref{table_fovs}), twice as high as the
value measured by \citet{Lawrence93} and our AR peak contrast reaches 0.5 \% or larger (cf. Table
\ref{table_fovs}), whereas TTL's average contrasts in ARs were negative for all bins of magnetogram signal.

We checked that the dominant majority of the pixels distributed around the peak of the trends, i.e. with $B_{\rm
app,los}$ in an interval of $\pm 200$ G centered on the peak value (between dashed lines in Fig.
\ref{fig_C_vs_B_FOV}), are relatively well located in intergranular lanes (see contours in Fig. \ref{fig_FOV}),
as expected for magnetic elements, or oftentimes surround larger darker features. The probability
density function (PDF) of the \emph{intrinsic} field strength $B$ of these pixels reveals that they harbour kG
fields in the AR and in the QS, with nearly the same mean value $\left<B\right>$ of  approx. 1150 G (see Fig.
\ref{fig_PDFs_AR_QS}a). Since the peak of the contrast trends occurs at similar $B_{\rm app,los}$ for ARs and the
QS, the corresponding magnetic elements should also have similar filling factors $\alpha$ (unless their
inclination deviates significantly from the vertical), as verified by the PDF($\alpha$), which have mean values
$\left<\alpha\right>$ of 0.63 and 0.62 in the AR and in the QS, respectively (Fig. \ref{fig_PDFs_AR_QS}b).
Assuming that the filling factor scales with the size of the magnetic elements (see Sect. \ref{sec_concl}), this
in turn poses the problem of how to explain their different contrast in the QS and in ARs. As summarized
in Table \ref{table_fovs}, repeating the same analysis on different FOVs of QS and ARs yielded the same
conclusions.

\begin{table*}
\caption{List of the different fields of view (FOVs) analyzed in this work, extracted from the SP scans indicated
by their date (see Table \ref{table_scans}).} \vskip3mm \centering
\begin{tabular}{c c c c c c c c} 
\hline\hline
FOV index & date (dd-mm-yy) & target & Contrast peak & $B_{\rm app,los}$ peak [G] &  $\left<B\right>$ [G] & $\left<\alpha\right>$ &
$\left<\gamma\right>$ [$^\circ$]\\
\hline
1 & 11-12-06 & AR &  0.013 &  677 &     1104 &     0.64 &     15.9\\
2 & 05-01-07 & AR & 0.015 &     694 &      1128 &     0.64 &      17.2\\
3 & 05-01-07 & AR & 0.027  &  711 &    1125 &   0.64 &  14.7\\
4 & *01-02-07 & AR &  0.005 & 685 &  1142 & 0.63 & 20\\
5 & 01-02-07 & AR &   0.013 &     711 &      1147 &     0.64 &      17.5\\
6 & 28-02-07 & AR & 0.018 & 694 & 1136 & 0.63 & 16.1\\
7 & 01-05-07 & AR &  0.008 &      668 &     1119 &     0.62 &      17.1\\
8 & 01-05-07 & AR &   0.005 &      668 &      1106 &     0.62 &      16.5\\
9 & 11-05-07 & AR & 0.016 &     651 &      1111 &     0.62 &      18.7\\
\hline
   &  & & \textbf{0.013} & \textbf{681} &   \textbf{1124} &     \textbf{0.63} &      \textbf{17.4}\\
\hline
10 & 10-03-07 & QS & 0.056 & 750 & 1166 & 0.6 & 9.7\\
11 & 10-03-07 & QS & 0.029 & 709 & 1137 & 0.61 & 12.3\\
12 & 10-03-07 & QS & 0.037 & 678 & 1096 & 0.6 & 9.5\\
13 & 23-04-07 & QS & 0.013 & 660 & 1091 & 0.6 & 13.2\\
14 & 23-04-07 & QS & 0.046 & 805 & 1208 & 0.63 & 10.3\\
15 &*24-04-07 & QS & 0.029 & 702.5 & 1142 & 0.62 & 12.8 \\
16 & 24-04-07 & QS & 0.062 & 725 & 1174 & 0.6 & 13.5 \\
17 & 27-04-07 & QS & 0.04 &      711 &      1113 &     0.63 &      12.4\\
18 & 27-04-07 & QS & 0.029 &      702 &      1122 &     0.63 &      12.5\\
19 & 27-04-07 & QS & 0.033 & 727 & 1146 & 0.63 & 13.1\\
\hline
  &  & & \textbf{0.037} & \textbf{717} &   \textbf{1140} &     \textbf{0.62} &      \textbf{11.9}\\
\hline
\end{tabular}
\tablefoot{The AR and QS FOVs described in Sect. \ref{sec_CvsB} have their date marked by an asterisk. When
several FOVs are extracted from the same scan they share the same date. All these FOVs have sizes of $70\arsec
\times 60 \arsec$, except for the dates 11-12-06 and 05-01-07 for which the FOVs are $40\arsec \times 90 \arsec$
and $45\arsec \times 91 \arsec$, respectively. The ``Contrast peak'' refers to the maximum of the fit of
continuum contrast vs. $B_{\rm app,los}$ and ``$B_{\rm app,los}$ peak'' is the corresponding $B_{\rm app,los}$ at
the peak. $\left<B\right>$, $\left<\alpha\right>$, $\left<\gamma\right>$ are calculated over the pixels having
$B_{\rm app,los}$ in a $\pm 200$ G interval centered around the peak value of $B_{\rm app,los}$. The rows with
bold characters indicate the average values of the different quantities for the AR (middle row) and QS FOVs (last
row), respectively.}
\label{table_fovs}
\end{table*}

\subsection{Inclination of the magnetic elements}
\label{sec_incl}

Based on their observed center-to-limb variation of continuum contrast, \citet{Topka92} deduced that the contrast
of magnetic elements should be sensitive to the angle between the line-of-sight (LOS) and the magnetic lines of
force, which at disk center directly translates into their inclination with respect to the local vertical.
\citet{Lawrence93} then proposed that the larger contrasts in the quiet Sun could be explained by a larger
inclination of the magnetic elements, whereby their hot walls would be more visible.

\begin{figure}
\centering
\includegraphics[width=0.9\columnwidth]{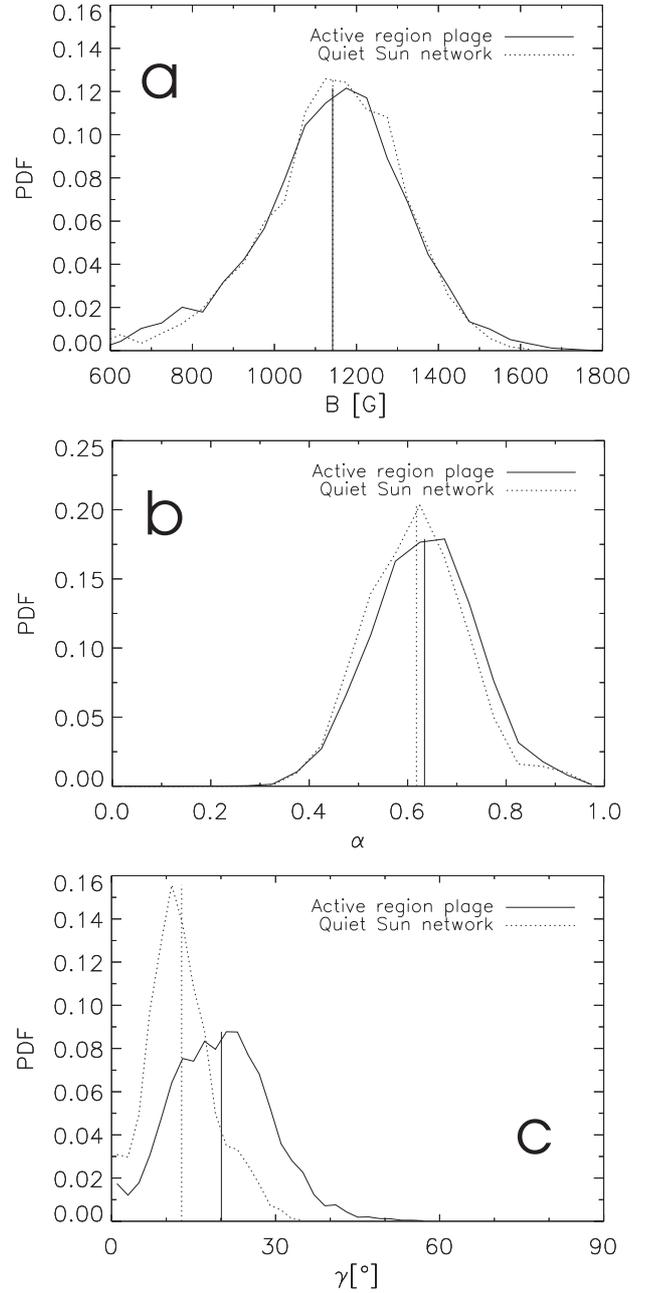}
\caption{Probability density functions (PDFs) of (a) the intrinsic field strength, $B$, (b) the magnetic filling
factor, $\alpha$ and (c) the inclination $\gamma$ of the pixels having $B_{\rm app,los}$ in an interval of $\pm
200$ G around the peak value of $B_{\rm app,los}$, for the quiet Sun (dotted curves) and the active
region (solid curves) fields of view shown in Fig. \ref{fig_FOV}. The vertical lines mark the mean values of the
distributions.}
\label{fig_PDFs_AR_QS}
\end{figure}

We have ruled out this possibility by studying the probability density functions (PDFs) of the inclination
$\gamma$ for the pixels identified with magnetic elements having $B_{\rm app,los}$ in a $\pm 200$ G interval
around the peak value of the contrast curves (between dashed lines in Fig. \ref{fig_C_vs_B_FOV}). Note that in
this interval of $B_{\rm app,los}$, all the pixels have a Stokes $V$ amplitude larger than 30 times the rms noise
level, while 40 \% of these pixels also have $Q,U$ amplitudes above 4.5 times their noise level. Figure
\ref{fig_PDFs_AR_QS}c displays the PDF($\gamma$) for the AR plage and QS FOVs shown in Figs
\ref{fig_FOV}.\footnote{As we are here only interested in quantifing the deviation from the vertical, all the
inclinations $\gamma$ are reported in the range $0^\circ < \gamma < 90^\circ$.} The magnetic elements are close
to vertical in both cases, with a distinctly larger $\gamma$ in the AR ,$\left< \gamma \right > = 19.9 \pm 0.1
^\circ$ (assuming Gaussian statistics), than in the QS, $\left< \gamma \right > = 12.8 \pm 0.1 ^\circ$. Probing
different FOVs yielded similar PDFs and mean inclinations (see Table \ref{table_fovs}). Hence, the argument of
\citet{Lawrence93} does not hold. Instead of larger inclinations in the QS, the mean inclinations are
systematically (although marginally) larger by 3$^\circ$ to 10$^\circ$ in the AR FOVs. The similarity in the PDFs
of $B$ and $\alpha$ strongly argues against a bias responsible for the different inclinations. A check at the
inclination maps revealed that these larger inclinations in the ARs mostly occur at the periphery of pores and of
dense groups of magnetic features typical of plages, while more isolated magnetic elements like in the QS tend to
be more vertical. This is expected as larger flux concentrations like pores have larger vertical expansion rates,
which tends to bend the file lines of neighboring magnetic elements. Likewise, a magnetic element located at the
periphery of a closely packed group of magnetic features feels a net bending force towards the outside of the
group.

Hence, up to the small difference in inclination discussed above, from Fig. \ref{fig_PDFs_AR_QS} and Table
\ref{table_fovs} we can conclude that all quantities ($B, \alpha, \gamma$) have rather similar values for the
brightest magnetic elements in the QS and in ARs, which still leaves unexplained the fact that they reach larger
contrasts in the QS compared to ARs.

\subsection{Positive and negative polarity fields in ARs}
\label{sec_polarity}

So far, we have only dealt with the \emph{strength} of the longitudinal magnetic field without considering the
different polarities in the fields of view (FOVs), mainly because our AR FOVs are generally dominated by
a single polarity.

However, it is possible that in ARs the contrast of the two opposite polarities differ due to their possible
interaction and/or different inclination of their magnetic elements. For instance, cases of Moving Magnetic
Features (MMFs) in a Sunspot moat have been reported to interact with the opposite polarity magnetic elements
yielding chromospheric surges in H$\alpha$ \citep{Brooks07}. Such interacting features could exhibit different
contrasts, possibly related to different inclinations \citep[following the argument of][see Sect.
\ref{sec_incl}]{Topka92}. When analyzing the continuum contrast vs. magnetogram signal separately of the two
polarities, \citet{Topka92} found a difference between the contrasts reached by the two polarities (with either
the positive or negative polarity being brighter depending on the FOV), for different instances of AR plages
located at heliocentric angles $\theta = 4,7$ and $43^\circ$. They concluded that this
angle had to be different between the two polarities \citep[employing the same reasoning as][to explain the different
contrasts between QS network and AR plage, see Sect. \ref{sec_incl}]{Lawrence93}.

\begin{figure*}
\centering
\includegraphics[width=0.8\textwidth]{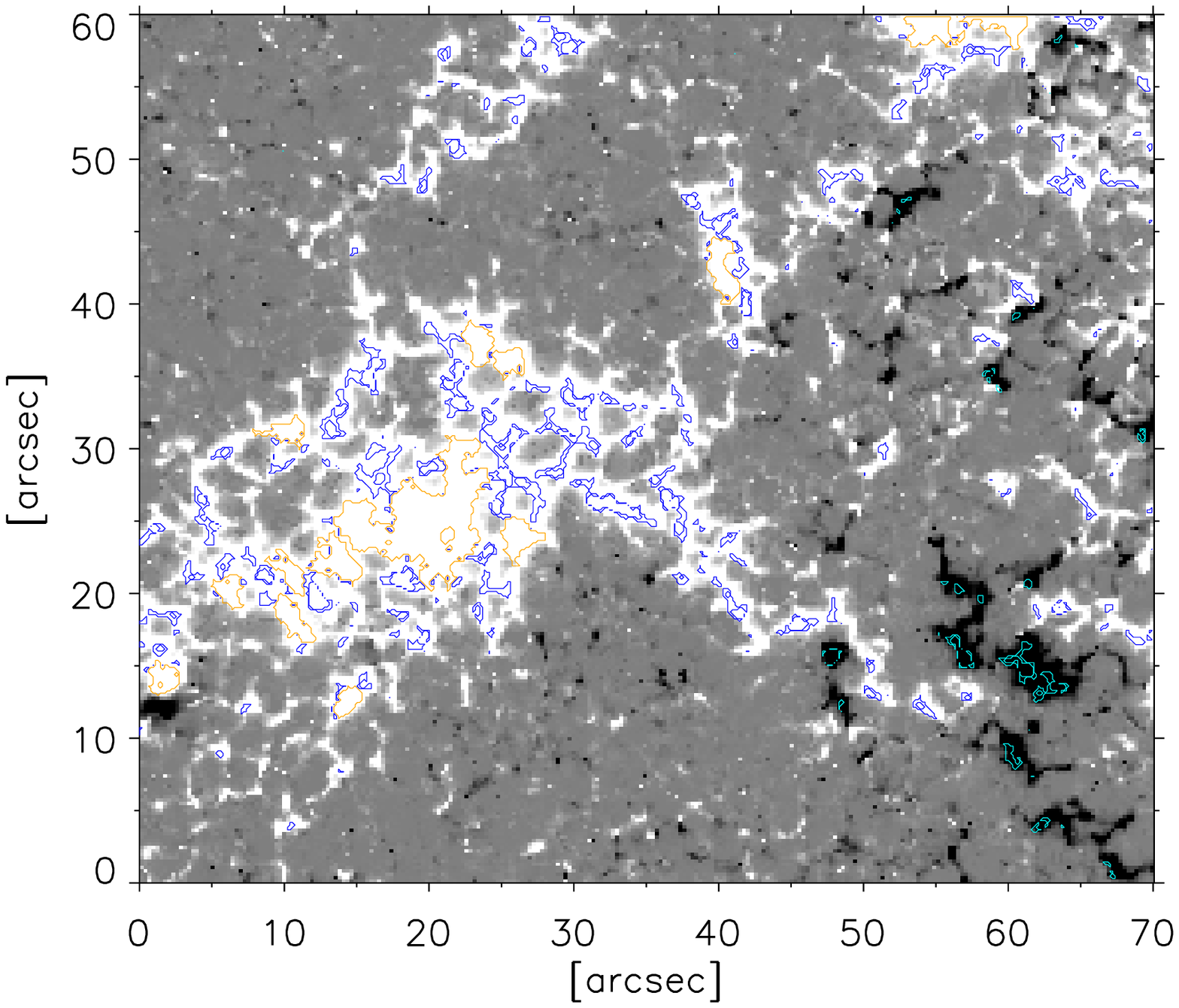}
\caption{Map of ``signed $B_{\rm app,los}$'' for the AR FOV extracted from the SP scan dated 01-02-07 (FOV index
5, see Table \ref{table_fovs}. The map is saturated at $+100$ G (white) and $-100$ G (black). The pores
eliminated from the contrast analysis are contoured in orange (there is no pores in the negative polarity). The
pixels having $B_{\rm app,los}$ in an interval of $\pm 200$ G around the contrast peaks of each polarity are
contoured in blue (positive polarity) and cyan (negative polarity).}
\label{fig_AR_B_polarity}
\end{figure*}

\begin{figure*}
\centering
\includegraphics[width=0.9\textwidth]{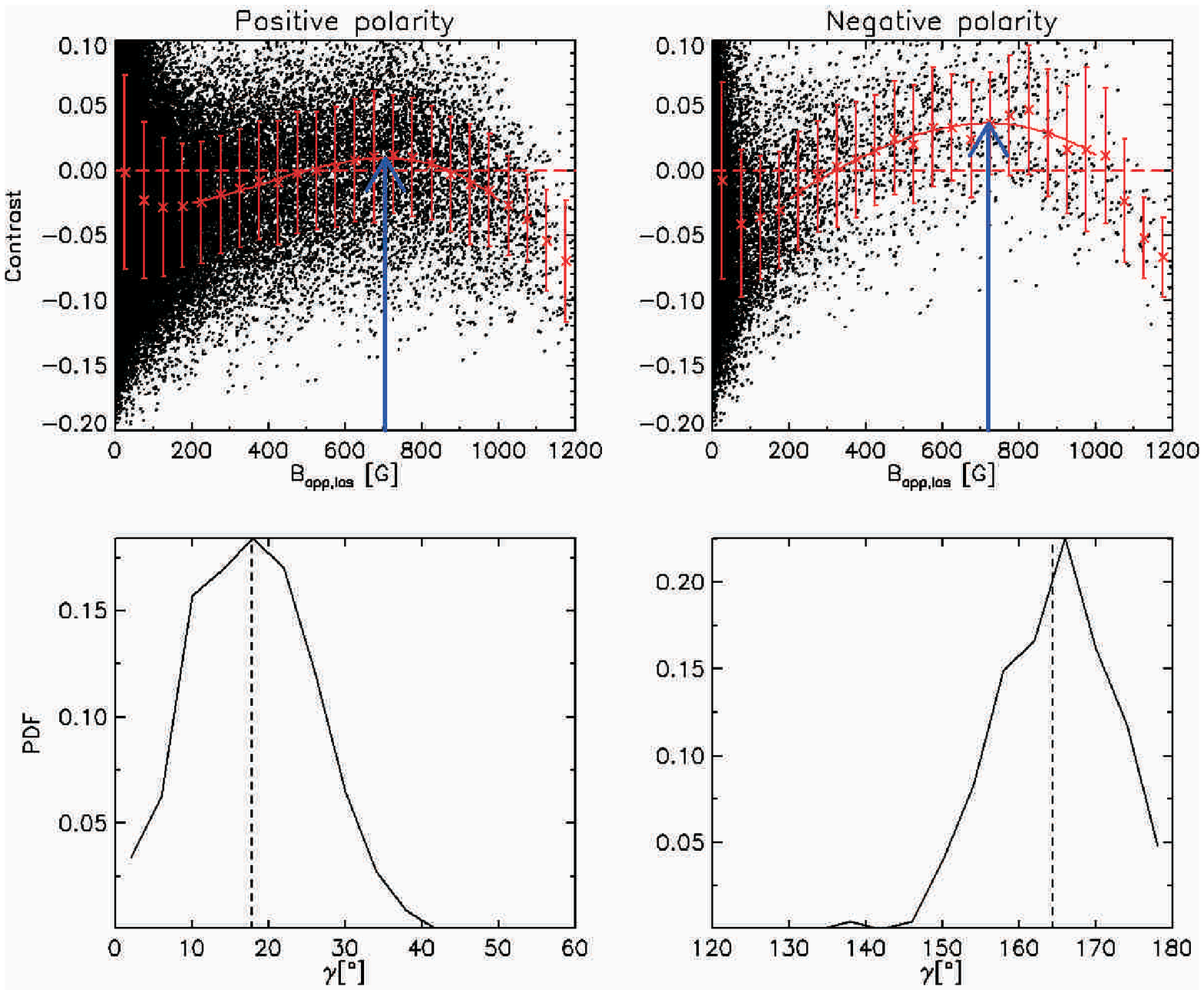}
\caption{Upper panels: Scatterplot of continuum contrast vs. $B_{\rm app,los}$ for the positive and
negative polarity in the FOV displayed in Fig. \ref{fig_AR_B_polarity}. Symbols are as in Fig.
\ref{fig_C_vs_B_FOV}.
Lower panels: Probability distribution functions of the inclination $\gamma$ retrieved by the inversion for the
positive (between 0$^\circ$ and 90$^\circ$) and negative (between 90$^\circ$ and 180$^\circ$) polarity.}
\label{fig_AR_polarity}
\end{figure*}

In six cases, our ARs FOVs contained enough flux of both polarities to investigate their behaviour separately. We
present here one example, while we list the results obtained from the other five FOVs in Table
\ref{table_polarity}. Figure \ref{fig_AR_B_polarity} shows a map of the ``signed $B_{\rm app,los}$'' (i.e. the
longitudinal component of the apparent magnetic field) saturated at $\pm 100$ G, to illustrate the distribution
of the patches of both polarity fields in the example FOV (positive in white, negative in black). In fact, the
minority (negative) polarity is essentially distributed close to the Sunspot (on the immediate right out of the
FOV) and probably corresponds to footpoints of magnetic field lines from the penumbra.
Repeating the pixel-by-pixel analysis of contrast vs. $B_{\rm app,los}$ for both polarities as
in Fig. \ref{fig_AR_polarity} reveals that, although the contrasts peak at $B_{\rm app,los} \sim 700$ G for both
polarities, the minority polarity reaches larger contrasts by almost 3\%. However, the probability
distribution function of the inclination $\gamma$ for the pixels with $B_{\rm app,los}$ in an interval of $\pm
200$ G around the contrast peaks of each polarity (blue and cyan contours in Fig.
\ref{fig_AR_B_polarity}, associated with the magnetic elements producing the contrast peak) does not show any
noticeable difference between both polarities, as visible in Fig.
\ref{fig_AR_polarity} (lower panels).
As one can see in Table \ref{table_polarity}, in none of the six FOVs does the mean inclination
$\left<\gamma\right>_{\pm}$ of the magnetic elements differ between the positive ($+$) and negative ($-$)
polarity, although the peak contrasts differ by 0.1 to 3.9 \%. Thus, the observed contrast difference between
polarities can not be accounted for by different inclinations.

We noticed that the peak contrast was always larger in the minority polarity and that in general, the
contrast difference between polarities was larger for FOVs in which the polarities were seemingly more
``unbalanced'' (i.e. the ratio of surface coverage between majority and minority polarity was larger). To
quantify this,  we calculated the fraction of the total unsigned flux carried by each polarity as:
\begin{equation}
X_{\pm} = \frac{\int_{\rm FOV_{\pm}} {\rm d} A_{\pm} B_{\rm app,los}}{\int_{\rm FOV} {\rm d} A B_{\rm app,los}},
\end{equation}
where the integration $\int_{\rm FOV_{\pm}}$ is performed over those areas covered by the positive or negative
polarity. Note that pores were included in this calculation.
In the three cases where the majority polarity carries $>70$\% of the flux (labeled by superscript ``a'' in
Table \ref{table_polarity}), the peak contrast of the majority polarity is systematically significantly lower than
that of the minority polarity by 1.3 to 3.9 \%. In one exception (labeled ``b''), although the
majority polarity contains twice as much flux as the minority polarity, the peak contrast of the majority
polarity is only marginally lower. For the remaining two cases where the fluxes were rather balanced (labeled
``c''), the peak contrasts are very similar in one case but differ by about 3\% in the other one (FOV index 9 in
the Table). This specific FOV differs from all the others in that it harbours the only small bipolar ephemeral
region in our dataset, while all the other FOVs contain plage areas in the vicinity of sunspots. Moreover, a look at the full
scan containing this FOV reveals that the positive polarity is in fact by far dominating, but that most of its
flux is located outside of the selected FOV (the rest of the area was not selected since it lays at $\mu < 0.99$).

We propose that the lower contrast of the predominant polarity is explained by the larger size of its magnetic
field patches (saturated white in Fig. \ref{fig_AR_B_polarity}). These represent larger obstacles to the
convective flows and thereby are prone to contain cooler magnetic elements, whereas the patches of the minority
polarity are rather reminiscnent of the QS network. The influence of such magnetic patches on the surrounding
convection will be presented in more details in a forthcoming paper \citep[][Paper II of this
series]{Kobel11b}.

\begin{table*}
\caption{Values of peak contrast, mean inclination $\left<\gamma\right>_{\pm}$ and fraction of the total unsigned
flux $X_{\pm}$ in the positive ($+$) and negative ($-$) polarities for the AR FOVs analyzed in Sect. \ref{sec_polarity}.}
\vskip3mm \centering
\begin{tabular}{c c c c c c c c} 
\hline\hline
FOV index & Contrast peak ($+)$ & Contrast peak ($-$) &  $\left<\gamma\right>_{+}$ [$^\circ$] & $\left<\gamma\right>_{-}$ [$^\circ$] &
$X_{+}$ & $X_{-}$ \\
\hline
*5\tablefootmark{a} & 0.009 & 0.036 & 17.8 & 164.5 & 90.5 & 9.5 \\
6\tablefootmark{a} & 0.015 & 0.054 & 16 & 166.7 & 89.5 & 10.5 \\
1\tablefootmark{a} & 0.024 & 0.011 & 16.4 & 164.5 & 27.7 & 72.3 \\
2\tablefootmark{b} & 0.017 & 0.013 & 17.2 & 162.4 & 32.3 & 67.7 \\
3\tablefootmark{c} & 0.027 & 0.026 & 15.2 & 166 & 53.5 & 46.5 \\
9\tablefootmark{c} & 0 & 0.029 & 18 & 160 & 53.6 & 46.4\\
\hline
\end{tabular}
\label{table_polarity}
\tablefoot{The FOV index is the same as in Table
\ref{table_fovs}.
The example FOV presented in Figs. \ref{fig_AR_B_polarity} and \ref{fig_AR_polarity} has and asterisk before its
index.
$\left<\gamma\right>_{\pm}$ was calculated only for the pixels distributed in an interval of
$\pm200$ G around the $B_{\rm app,los}$ location of the contrast peak in each polarity.}
\end{table*}

\subsection{Intrinsic brightness of the magnetic elements}
\label{sec_brightness}

One possibility to explain the lower contrast of magnetic elements in ARs compared to the QS is that the absolute
brightness of the chosen contrast reference $\left< I_c \right>_{\rm ref, FOV}$ (mean continuum intensity of the
pixels having $B_{\rm app,los} < 100$ G in the FOV, see Eq. \ref{eq_contrast}) is systematically larger in ARs
compared to the QS, such as to account for the relative difference of 2.4 \% in the average peak contrasts (see
Sect. \ref{sec_CvsB} and Table \ref{table_fovs}). This hypothesis is a priori supported by the idea of global
horizontal convective inflows developing towards ARs to compensate for the radiative losses through the magnetic
atmosphere \citep{Komm93, Spruit03, Zhao04, Rempel06}, in a similar fashion as in the case of the case of simple
2D flux sheets \citep{Deinzer84}. A possible opacity reduction of the magnetic atmosphere above granulation
near plages could also contribute to brighter contrast references in the AR FOVs.

\begin{figure}
\centering
\includegraphics[width=\columnwidth]{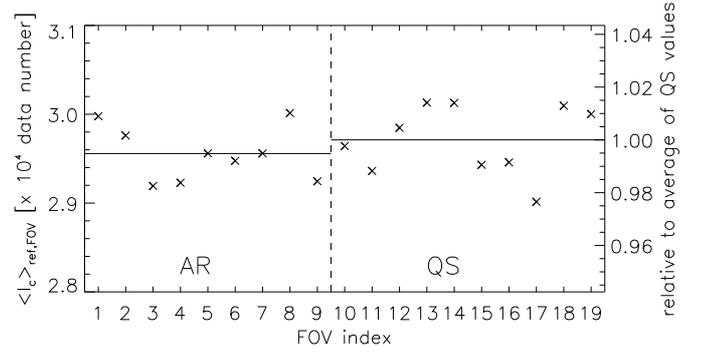}
\caption{Values of the contrast references $\left< I_c \right>_{\rm ref, FOV}$ used in the different AR and QS
FOVs, in original data units. The FOV index labels the FOVs according to the order of Table \ref{table_fovs}.
The horizontal bars represent the average values of the $\left< I_c \right>_{\rm ref, FOV}$ in the AR (left) and QS (right) FOVs.
The right-hand side y axis gives the values relative to the QS average.}
\label{fig_refFOV}
\end{figure}

To check this possibility, we compared the (intensity) ``instrumental data number'' values (i.e. those of
level 1 data obtained after calibration with \verb"sp_prep"\footnote{For terminology, see the Hinode SOT Data
Analysis Guide at \url{http://solarwww.mtk.nao.ac.jp/katsukaw/sot_fits/SOT00042_C_SOT_Analysis_Guide_(SAG).pdf}.}, cf. Sect.
\ref{sec_Ic_mu}) of the contrast references $\left< I_c \right>_{\rm ref, FOV}$ used in the different FOVs, as
plotted in Fig. \ref{fig_refFOV}. In addition to the effect mentioned above, the $\left< I_c \right>_{\rm ref,
FOV}$ values can also vary due to oscillation-induced brightness fluctuations, possible differences in convection
and instrumental variations from one scan to the next \citep[for instance, varying amount of defocus,
see][]{Danilovic08}. All this leads to a scatter of the values with a rms of 1.2\%. To look for a systematic
difference between the QS and AR contrast references, we averaged those values and overplotted the averages of
the QS and the AR $\left< I_c \right>_{\rm ref, FOV}$ as horizontal bars. As indicated by the right-hand side
y-axis of the plot, the average of the $\left< I_c \right>_{\rm ref, FOV}$ of the AR FOVs is lower than the
average value of the QS FOVs and differs by only $-0.5 \pm 0.2$ \% (the error is the sum of the errors on the QS
and AR averages). This is five times smaller and of opposite sign than the relative difference of 2.4 \% between
the average peak contrasts in ARs and QS. We can thus rule out that the lower observed contrasts of magnetic
elements in ARs compared to the QS are related to systematically brighter contrast reference areas in ARs.

\subsection{Comparison with previous studies of contrast vs. magnetogram signal}
\label{sec_comp}

Here we propose possible explanations as to why our trends of continuum contrast averaged in bins of apparent
longitudinal field strength $B_{\rm app, los}$ peak at finite $B_{\rm app, los}$ in ARs, whereas the
trends of \citet{Title92, Topka92, Lawrence93} (TTL) in ARs decrease monotonically.

\begin{figure*}
\centering
\includegraphics[width=\textwidth]{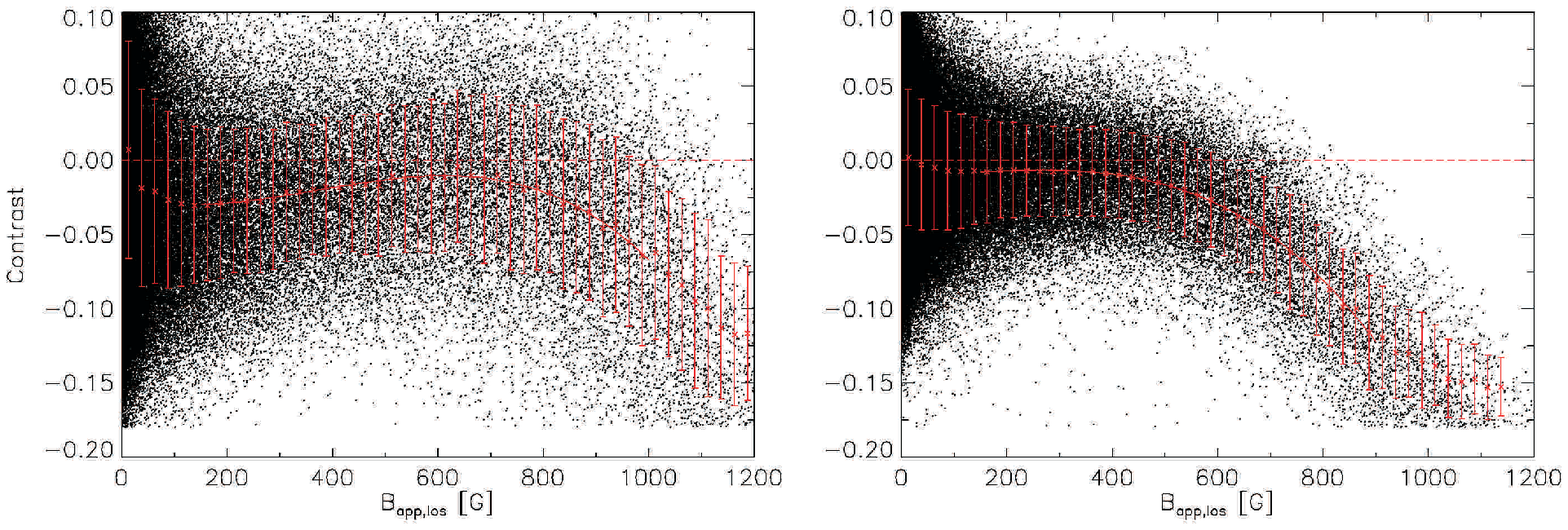}
\caption{left: Scatterplot of the continuum contrast vs. longitudinal flux density $B_{\rm app,los}$ for the
plage field of view (FOV) shown in Fig. \ref{fig_FOV}, obtained by removing the pores via a simple contrast
threshold at a value of $-0.18$. right: Scatterplot of the continuum contrast vs. apparent longitudinal field
strength $B_{\rm app,los}$ for the same plage FOV, obtained after degradation of the contrast and $B_{\rm
app,los}$ by convolving with a Gaussian of FWHM $0\dotarsec45$ and a Lorentzian of FWHM $0\dotarsec06$ (mimicing
straylight), while removing the pores with the same contrast threshold.}
\label{fig_C_vs_B_deg}
\end{figure*}

\begin{figure*}
\centering
\includegraphics[width=0.8\textwidth]{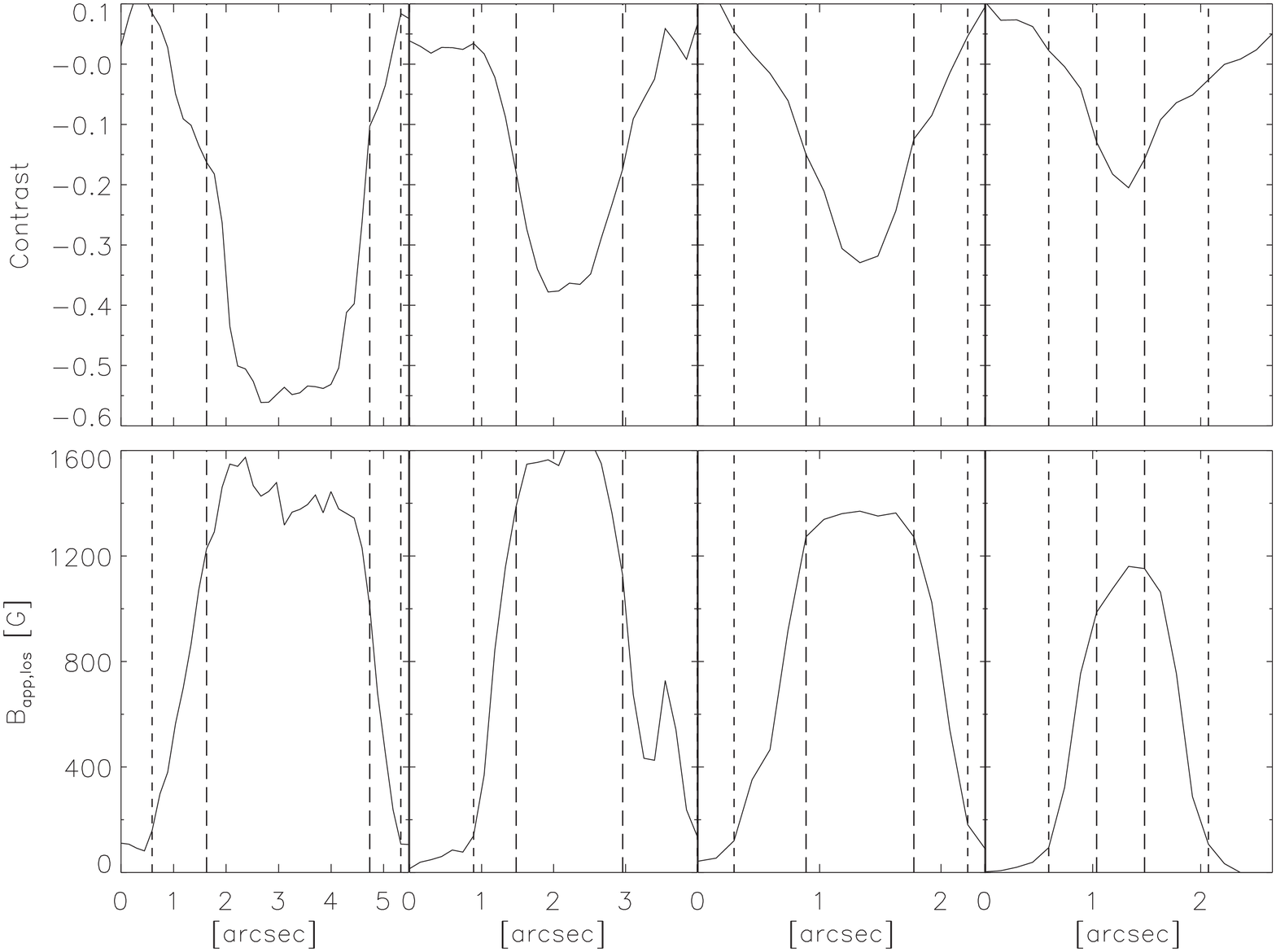}
\caption{Contrast (upper panels) and $B_{\rm app,los}$ (lower panels) cuts across four pores of decreasing size.
The pixels located between the long dashed lines have contrasts inferior to $-0.18$, and the short dashed lines
enclose the pixels having $B_{\rm app,los} > 200$ G. Note that due to the pixellation of the profiles, the
vertical dashed lines do not cross them exactly at those values. All cuts have been extracted from the plage
field of view shown in Fig. \ref{fig_FOV} and their locations are indicated in that Figure by white solid lines.}
\label{fig_cutpores}
\end{figure*}

One source of discrepancy lies in the different pore removal techniques employed by TTL and by us. If instead of our pore
removal procedure, described in Sect. \ref{sec_CvsB}, we use a simple contrast threshold as employed by TTL on
the plage FOV presented in Fig. \ref{fig_FOV}, the resulting trend of the contrast vs. $B_{\rm app,los}$ is much
flatter. For instance, by cutting out all the pixels with contrast below $-0.18$, we obtain the scatterplot
displayed in Fig. \ref{fig_C_vs_B_deg} left panel, while our original pore removal leads to Fig.
\ref{fig_C_vs_B_FOV} lower panel.\footnote{Note that due to different spatial resolutions and contrasts, we
cannot use the same threshold values as TTL. The threshold value of $-0.18$ used here was chosen to be as high as
possible without cutitng in intergranular lanes.} The peak of the trend at 600-700 G is greatly reduced in Fig.
\ref{fig_C_vs_B_deg} (the peak contrast reaches only $-0.01$ compared to $0.005$ in Fig. \ref{fig_C_vs_B_FOV}),
while the darkening at $B_{\rm app,los} > 800$ G is significantly enhanced. This is because a simple intensity
cut only removes the inner dark cores of the pores, which mostly contribute to the range $B_{\rm app,los} > 800$
G. But our more-complex procedure also removes the immediate surroundings of the pores, where the pixels are
darkened by the point spread function of the telescope. Such pixels typically provide contributions in the range
200-800 G and if not removed, tend to flatten the peak of the contrast curves. The effect of the telescope
diffraction on the surroundings of the pores is illustrated by cuts of contrast and $B_{\rm app,los}$ across
pores of different sizes, as plotted in Fig. ~\ref{fig_cutpores}. The cuts were extracted from the plage area of
Fig. ~\ref{fig_FOV} where they are marked by thick white lines. On both sides of the dark cores (identified in
the cuts as having contrast $< -0.18$ and marked by the long dashed lines in Fig. \ref{fig_cutpores}), we find
several pixels where the contrast is slightly negative or neutral and $B_{\rm app,los} > 200$ G. To remove these
pixels, our pore removal procedure extends the detected dark cores of pores (which we actually require to have
contrasts $< -0.15$ \emph{and} $B_{\rm app,los} > 900$ G, see Sect. \ref{sec_CvsB}) until $B_{\rm app,los}$ drops
below 200 G. The boundaries obtained in this manner are marked by short-dashed lines in Fig. \ref{fig_cutpores}.

To reproduce a monotonically decreasing contrast curve like the ones of TTL, we also had to degrade our data.
These authors indeed state that the spatial resolution is at best $0\dotarsec3$ and $0\dotarsec45$ for their
images and magnetograms, respectively. We obtained a good match to TTL's results if we degraded \emph{both} the
continuum images and Stokes images with a Gaussian of $0\dotarsec45$ FWHM, plus an additional Lorentzian
degradation of $0\dotarsec06$ FWHM to mimic straylight. This yielded the monotonically decreasing contrast trend
of Fig. \ref{fig_C_vs_B_deg} right, provided the pores were eliminated via a simple intensity cut as described
before.\footnote{If before the degradation we eliminate the pores with our own pore removal procedure, the
resulting curve (not shown here) still exhibits a peak.} Although the Gaussian degradation already reduces the
contrast range to values similar to TTL's, the additional Lorentzian degradation was needed in order to obtain
such a monotonous decrease. The width of that Lorentzian was adjusted so that the rms contrast calculated in a
$20\arsec$ box of quiet Sun in our images matches the one given by \citet{Topka92} (6\% at 558 nm, corresponding
to about 5\% at 630 nm). This additional degradation implies that TTL's results are probably sgnificantly influenced by
straylight, which is very typical for ground-based observations. In contrast, the small amount of scattered light
in Hinode's spectropolarimeter \citep{Danilovic08} can be neglected. However, we warn against inferring that the
FWHM of our degradations corresponds to the actual resolution of TTL. As will be shown in a second paper
\citep[][in prep.]{Kobel11b}, the shape and peak value of the contrast vs. $B_{\rm app,los}$ trend also depend on
the amount of flux contained in the FOV.

That the spatial resolution is an important factor influencing the peak of the contrast curves can also
be seen by comparing with the more recent studies of \citet{Berger07} and \citet{Narayan10}, using higher
resolution ($0\dotarsec15$) SST data consisting in 630.2 nm magnetograms, together with G-band and 630.2 nm
continuum filtergrams, respectively (exclusively in ARs). Using the same pixel-by-pixel method as us, both groups
of authors found a peak contrast reaching up to 2-3 \% (of the mean intensity in the FOV) in the study of
\citet{Narayan10} and 3-4 \% in the case of \citet{Berger07}. The latter study, however, is not directly
comparable since magnetic elements are known to exhibit enhanced contrasts in G-band \citep[e.g.][]{Muller84}.
Future studies at higher resolution and lower straylight thus promise to reach even higher contrasts \citep[e.g.
with the SUNRISE balloon-borne observatory][]{SUNRISE_mis, SUNRISE_res, Riethmueller10}. Note that in spite of
his much lower spatial resolution (magnetograph aperture of $2\dotarsec4$), \citet{Frazier71} did find peaked
contrast curves in two active regions. According to the above, we can possibly attribute this to a lower amount
of straylight (compared to TTL), the presence of only a few pores in his field of view, or a lower degree of
magnetic activity.

\section{Discussion and conclusion}
\label{sec_concl}

Owing to its better spatial resolution and its low amount of scattered light, Hinode/SP allowed us to update the
results of earlier ground-based studies of continuum contrast vs. magnetogram signal \citep[][referred to in
this paper as TTL]{Title92, Topka92, Topka97, Lawrence93} in the QS and in ARs.
In agreement with the previous finding of \citet{Lawrence93}, the contrasts reach larger values in the QS than in
ARs, with a relative contrast difference of 2.5\% (average over 8 and 10 fields of view in ARs and the QS,
respectively). This difference could not be attributed to any brightness enhancement of the reference areas in
ARs (see Sect. \ref{sec_brightness}), so that it likely relates to an intrinsic brightness difference between the
brightest magnetic elements in ARs and in the QS.
Whereas TTL were finding monotonically decreasing trends in ARs, the trends of the contrast vs. (apparent)
longitudinal field strength $B_{\rm app,los}$ now display a clear peak \emph{both} in ARs and in the QS, as
visible in Fig. \ref{fig_C_vs_B_deg}. We explained our peak in ARs by the higher spatial resolution and low
amount of straylight of Hinode compared with the earlier ground-based observations of TTL, as well as by our removal of the
pores and the part of their close surroundings affected by the telescope diffraction (tending to smear the
contrast curve). This contrast peak enables us to attribute an apparent longitudinal field strength value to
the brightest pixels in ARs and QS. Interestingly, the apparent
longitudinal field strength corresponding to the peak contrast is similar for ARs and QS, $B_{\rm app,los} \sim
700$ G.
Note that the continuum contrast curve of \citet[][Fig. 2 upper right panel]{Narayan10} also peaks at
$B_{\rm app,los} \sim 700$ G, although they used data from the Swedish Solar Telescope (SST). Given the
theoretically factor of 2 higher spatial resolution of SST compared to Hinode, this is surprising in the light of
recent tests with degraded MHD simulation snapshots. These indicate that the presence of a peak as well as its
associated $B_{\rm app,los}$ value depend on the spatial resolution \citep{Roehrbein11}.\footnote{For original
snapshots of simulations with $\left<B\right>$ up to 200 G, the bolometric intensity increases with B \citep[see
also ][]{Voegler05} but shows a peak after degradation.} However, as mentioned by \citet{Narayan10}, it is
likely that their $B_{\rm app,los}$ are underestimated due to instrumental and seeing-induced straylight. This
may explain the similar $B_{\rm app,los}$ of the peak. In this respect, it would be interesting to determine the
contrasts of BPs in ARs at still higher spatial resolution, as afforded by SUNRISE \citep{SUNRISE_mis}, should
appropriate data become available from a future flight.

We checked that the pixels contributing to the contrast peak (i.e. with $B_{\rm app,los}$ around 700 G) are
mainly located in intergranular lanes (cf. Fig. \ref{fig_FOV}) and can thus be reasonably associated with
magnetic elements. From the inversions we found that for these pixels the intrinsic magnetic field is kG in
strength and rather vertical (as expected for magnetic elements), as shown in Fig. \ref{fig_PDFs_AR_QS}a,b.
Consequently, the magnetic elements with higher continuum contrasts also share similar filling factors in ARs and
in the QS (see Fig. \ref{fig_PDFs_AR_QS}c), which poses the problem of how to interpret their larger brightness in
the QS.

It can be argued that at Hinode's spatial resolution it is unlikely that many flux tubes are present in a pixel,
and consequently that the filling factor in most cases reflects the size of the (unresolved) magnetic elements.
This is supported by recent results obtained with SUNRISE/IMaX \citep[spatial resolution between $0\dotarsec15$
and $0\dotarsec18$,][]{Imax10} concluding that many of the QS magnetic elements were spatially resolved
\citep{Lagg10} even in the internetwork. Since the resolution of Hinode/SP is only a factor of 2 inferior ($\sim
0\dotarsec3$), it must be close to resolving them, although we cannot rule out that even smaller kG features are
present. Under the assumption of a single magnetic elements per pixel on average, one could deduce that the brightest
magnetic elements have similar sizes in ARs and in the QS in spite of their different brightness, which in turn
questions the conventional interpretation that the brightness of magnetic elements is primarily dictated by their
size.

The only inversion parameter studied here that slightly differs between the brightest magnetic elements in ARs
and in the QS is the inclination, which is on average 5$^\circ$ larger in ARs (see Fig. \ref{fig_PDFs_AR_QS}c
and Table \ref{table_fovs}). However, we cannot think of any physical argument explaining why more inclined
magnetic elements appear darker. Rather, inclined flux tubes should appear brighter as their hot walls is more
exposed to a vertical line-of-sight \citep[as proposed by][]{Topka92}. We think that the more inclined fields
in ARs are just a consequence of the presence of pores and dense groups of flux tubes exerting a bending force
on the field lines of magnetic elements located at their periphery (due to the solenoidality of the magnetic
field and the expansion of the flux tubes with height).

Another factor that influences the heating of the magnetic elements is the efficiency of the surrounding
convection. It is well-known that the number density of magnetic elements (and thereby the average filling
factor) is higher in AR plages than in the QS network, and the ``degree of packing'' of magnetic elements is
likely to affect the surrounding convective motions \citep[as first suggested by][]{Knoelker88}.
\citet{Morinaga08} recently found (using Hinode observations of a pore and its surrounding plage) that the
disperion in line-of-sight velocity was smaller in regions of larger averaged filling factor, while
\citet{Ishikawa08} also observed reduced convective velocities in regions of high apparent longitudinal field.
This could also explain why the minority polarity magnetic elements in ARs, which are less densely packed and
form smaller patches, reach larger contrasts (in the few cases investigated in Sect. \ref{sec_polarity}). A
systematic study of whether the degree of inhibition of convection relates to the brightness of magnetic elements
will be presented in a second paper \citep[][in prep.]{Kobel11b}.

\begin{acknowledgements}
We gratefully thank M. Sch\"{u}ssler, D. R\"{o}rbein, B. Viticchi\'{e}, Y. Unruh, J. Sanchez-Almeida and the
ground-based solar observation group at the Max-Planck Institut f\"{u}r Sonnensystemforschung for interesting
discussions about this work. This work has been partially supported by WCU grant No. R31-10016 funded by the
Korean Ministry of Education Science and Technology. Hinode is a Japanese mission developed and launched by
ISAS/JAXA, collaborating with NAOJ, NASA and STFC (UK). Scientific operation of the Hinode mission is conducted
by the Hinode science team organized at ISAS/JAXA. This team mainly consists of scientists from institutes in the
partner countries. Support for the post-launch operation is provided by JAXA and NAOJ (Japan), STFC (U.K.), NASA,
ESA, and NSC (Norway). This work has also made use of the NASA ADS database.
\end{acknowledgements}

\bibliographystyle{aa}
\bibliography{biblio}

\begin{thebibliography}{83}
\expandafter\ifx\csname natexlab\endcsname\relax\def\natexlab#1{#1}\fi

\bibitem[{{Barthol} {et~al.}(2011){Barthol}, {Gandorfer}, {Solanki},
  {Sch{\"u}ssler}, {Chares}, {Curdt}, {Deutsch}, {Feller}, {Germerott},
  {Grauf}, {Heerlein}, {Hirzberger}, {Kolleck}, {Meller}, {M{\"u}ller},
  {Riethm{\"u}ller}, {Tomasch}, {Kn{\"o}lker}, {Lites}, {Card}, {Elmore},
  {Fox}, {Lecinski}, {Nelson}, {Summers}, {Watt}, {Mart{\'{\i}}nez Pillet},
  {Bonet}, {Schmidt}, {Berkefeld}, {Title}, {Domingo}, {Gasent Blesa}, {Del
  Toro Iniesta}, {L{\'o}pez Jim{\'e}nez}, {{\'A}lvarez-Herrero},
  {Sabau-Graziati}, {Widani}, {Haberler}, {H{\"a}rtel}, {Kampf}, {Levin},
  {P{\'e}rez Grande}, {Sanz-Andr{\'e}s}, \& {Schmidt}}]{SUNRISE_mis}
{Barthol}, P., {Gandorfer}, A., {Solanki}, S.~K., {et~al.} 2011, \solphys, 268,
  1

\bibitem[{{Berger} {et~al.}(1998){Berger}, {L{\"o}fdahl}, {Shine}, \&
  {Title}}]{Berger98}
{Berger}, T.~E., {L{\"o}fdahl}, M.~G., {Shine}, R.~S., \& {Title}, A.~M. 1998,
  \apj, 495, 973

\bibitem[{{Berger} {et~al.}(2007){Berger}, {Rouppe van der Voort}, \&
  {L{\"o}fdahl}}]{Berger07}
{Berger}, T.~E., {Rouppe van der Voort}, L., \& {L{\"o}fdahl}, M. 2007, \apj,
  661, 1272

\bibitem[{{Berger} \& {Title}(1996)}]{Berger96}
{Berger}, T.~E. \& {Title}, A.~M. 1996, \apj, 463, 365

\bibitem[{{Berger} \& {Title}(2001)}]{Berger01}
{Berger}, T.~E. \& {Title}, A.~M. 2001, \apj, 553, 449

\bibitem[{{Borrero} \& {Kobel}(2011)}]{BorreroKobel11}
{Borrero}, J.~M. \& {Kobel}, P. 2011, \aap, 527, A29+

\bibitem[{{Borrero} {et~al.}(2010){Borrero}, {Tomczyk}, {Kubo},
  {Socas-Navarro}, {Schou}, {Couvidat}, \& {Bogart}}]{Borrero10}
{Borrero}, J.~M., {Tomczyk}, S., {Kubo}, M., {et~al.} 2010, \solphys, 35

\bibitem[{{Brooks} {et~al.}(2007){Brooks}, {Kurokawa}, \& {Berger}}]{Brooks07}
{Brooks}, D.~H., {Kurokawa}, H., \& {Berger}, T.~E. 2007, \apj, 656, 1197

\bibitem[{{Danilovic} {et~al.}(2008){Danilovic}, {Gandorfer}, {Lagg},
  {Sch{\"u}ssler}, {Solanki}, {V{\"o}gler}, {Katsukawa}, \&
  {Tsuneta}}]{Danilovic08}
{Danilovic}, S., {Gandorfer}, A., {Lagg}, A., {et~al.} 2008, \aap, 484, L17

\bibitem[{{De Pontieu} {et~al.}(2006){De Pontieu}, {Carlsson}, {Stein}, {Rouppe
  van der Voort}, {L{\"o}fdahl}, {van Noort}, {Nordlund}, \&
  {Scharmer}}]{DePontieu06}
{De Pontieu}, B., {Carlsson}, M., {Stein}, R., {et~al.} 2006, \apj, 646, 1405

\bibitem[{{Deinzer} {et~al.}(1984){Deinzer}, {Hensler}, {Schussler}, \&
  {Weisshaar}}]{Deinzer84}
{Deinzer}, W., {Hensler}, G., {Schussler}, M., \& {Weisshaar}, E. 1984, \aap,
  139, 435

\bibitem[{{del Toro Iniesta}(2003)}]{Deltoro03}
{del Toro Iniesta}, J.~C. 2003, {Introduction to Spectropolarimetry}
  (Cambridge, UK: Cambridge University Press.)

\bibitem[{{Domingo} {et~al.}(2009){Domingo}, {Ermolli}, {Fox}, {Fr{\"o}hlich},
  {Haberreiter}, {Krivova}, {Kopp}, {Schmutz}, {Solanki}, {Spruit}, {Unruh}, \&
  {V{\"o}gler}}]{Domingo09}
{Domingo}, V., {Ermolli}, I., {Fox}, P., {et~al.} 2009, Space Science Reviews,
  145, 337

\bibitem[{{Frazier}(1971)}]{Frazier71}
{Frazier}, E.~N. 1971, \solphys, 21, 42

\bibitem[{{Frazier} \& {Stenflo}(1972)}]{Frazier72}
{Frazier}, E.~N. \& {Stenflo}, J.~O. 1972, \solphys, 27, 330

\bibitem[{{Hirzberger} \& {Wiehr}(2005)}]{Hirz05}
{Hirzberger}, J. \& {Wiehr}, E. 2005, \aap, 438, 1059

\bibitem[{{Howard} \& {Stenflo}(1972)}]{Howard72}
{Howard}, R. \& {Stenflo}, J.~O. 1972, \solphys, 22, 402

\bibitem[{{Ichimoto} {et~al.}(2008){Ichimoto}, {Lites}, {Elmore}, {Suematsu},
  {Tsuneta}, {Katsukawa}, {Shimizu}, {Shine}, {Tarbell}, {Title}, {Kiyohara},
  {Shinoda}, {Card}, {Lecinski}, {Streander}, {Nakagiri}, {Miyashita},
  {Noguchi}, {Hoffmann}, \& {Cruz}}]{HinodeSP2}
{Ichimoto}, K., {Lites}, B., {Elmore}, D., {et~al.} 2008, \solphys, 249, 233

\bibitem[{{Ishikawa} \& {Tsuneta}(2009)}]{Ishikawa09}
{Ishikawa}, R. \& {Tsuneta}, S. 2009, \aap, 495, 607

\bibitem[{{Ishikawa} {et~al.}(2008){Ishikawa}, {Tsuneta}, {Ichimoto}, {Isobe},
  {Katsukawa}, {Lites}, {Nagata}, {Shimizu}, {Shine}, {Suematsu}, {Tarbell}, \&
  {Title}}]{Ishikawa08}
{Ishikawa}, R., {Tsuneta}, S., {Ichimoto}, K., {et~al.} 2008, \aap, 481, L25

\bibitem[{{Keller} {et~al.}(2004){Keller}, {Sch{\"u}ssler}, {V{\"o}gler}, \&
  {Zakharov}}]{Keller04}
{Keller}, C.~U., {Sch{\"u}ssler}, M., {V{\"o}gler}, A., \& {Zakharov}, V. 2004,
  \apjl, 607, L59

\bibitem[{{Khomenko} {et~al.}(2003){Khomenko}, {Collados}, {Solanki}, {Lagg},
  \& {Trujillo Bueno}}]{Khomenko03}
{Khomenko}, E.~V., {Collados}, M., {Solanki}, S.~K., {Lagg}, A., \& {Trujillo
  Bueno}, J. 2003, \aap, 408, 1115

\bibitem[{{Kn\"{o}lker} {et~al.}(1991){Kn\"{o}lker}, {Grossmann-Doerth},
  {Sch\"{u}ssler}, \& {Weisshaar}}]{Knoelker91}
{Kn\"{o}lker}, M., {Grossmann-Doerth}, U., {Sch\"{u}ssler}, M., \& {Weisshaar},
  E. 1991, Adv. Space Res., 11, 285

\bibitem[{{Kn\"{o}lker} {et~al.}(1988){Kn\"{o}lker}, {Sch\"{u}ssler}, \&
  {Weisshaar}}]{Knoelker88}
{Kn\"{o}lker}, M., {Sch\"{u}ssler}, M., \& {Weisshaar}, E. 1988, \aap, 194, 257

\bibitem[{{Kobel} {et~al.}(2011{\natexlab{a}}){Kobel}, {Borrero}, \&
  J.~M.}]{Kobel11c}
{Kobel}, P., {Borrero}, \& J.~M., {Solanki}, S.~K. 2011{\natexlab{a}}, \aap,
  {in prep}

\bibitem[{{Kobel} {et~al.}(2011{\natexlab{b}}){Kobel}, {Borrero}, \&
  J.~M.}]{Kobel11b}
{Kobel}, P., {Borrero}, \& J.~M., {Solanki}, S.~K. 2011{\natexlab{b}}, \aap,
  {in prep}

\bibitem[{{Kobel} {et~al.}(2009){Kobel}, {Hirzberger}, {Solanki}, {Gandorfer},
  \& {Zakharov}}]{Kobel09}
{Kobel}, P., {Hirzberger}, J., {Solanki}, S.~K., {Gandorfer}, A., \&
  {Zakharov}, V. 2009, \aap, 502, 303

\bibitem[{{Komm} {et~al.}(1993){Komm}, {Howard}, \& {Harvey}}]{Komm93}
{Komm}, R.~W., {Howard}, R.~F., \& {Harvey}, J.~W. 1993, \solphys, 147, 207

\bibitem[{{Kosugi} {et~al.}(2007){Kosugi}, {Matsuzaki}, {Sakao}, {Shimizu},
  {Sone}, {Tachikawa}, {Hashimoto}, {Minesugi}, {Ohnishi}, {Yamada}, {Tsuneta},
  {Hara}, {Ichimoto}, {Suematsu}, {Shimojo}, {Watanabe}, {Shimada}, {Davis},
  {Hill}, {Owens}, {Title}, {Culhane}, {Harra}, {Doschek}, \&
  {Golub}}]{Hinode07}
{Kosugi}, T., {Matsuzaki}, K., {Sakao}, T., {et~al.} 2007, \solphys, 243, 3

\bibitem[{{Krivova} {et~al.}(2007){Krivova}, {Balmaceda}, \&
  {Solanki}}]{Krivova07}
{Krivova}, N.~A., {Balmaceda}, L., \& {Solanki}, S.~K. 2007, \aap, 467, 335

\bibitem[{{Krivova} {et~al.}(2003){Krivova}, {Solanki}, {Fligge}, \&
  {Unruh}}]{Krivova03}
{Krivova}, N.~A., {Solanki}, S.~K., {Fligge}, M., \& {Unruh}, Y.~C. 2003, \aap,
  399, L1

\bibitem[{{Krivova} {et~al.}(2006){Krivova}, {Solanki}, \& {Floyd}}]{Krivova06}
{Krivova}, N.~A., {Solanki}, S.~K., \& {Floyd}, L. 2006, \aap, 452, 631

\bibitem[{{Lagg} {et~al.}(2010){Lagg}, {Solanki}, {Riethm{\"u}ller},
  {Mart{\'{\i}}nez Pillet}, {Sch{\"u}ssler}, {Hirzberger}, {Feller}, {Borrero},
  {Schmidt}, {del Toro Iniesta}, {Bonet}, {Barthol}, {Berkefeld}, {Domingo},
  {Gandorfer}, {Kn{\"o}lker}, \& {Title}}]{Lagg10}
{Lagg}, A., {Solanki}, S.~K., {Riethm{\"u}ller}, T.~L., {et~al.} 2010, \apjl,
  723, L164

\bibitem[{{Landi Degl'Innocenti} \& {Landolfi}(2004)}]{Innocenti04}
{Landi Degl'Innocenti}, E. \& {Landolfi}, M., eds. 2004, Astrophysics and Space
  Science Library, Vol. 307, {Polarization in Spectral Lines}

\bibitem[{{Lawrence} {et~al.}(1993){Lawrence}, {Topka}, \&
  {Jones}}]{Lawrence93}
{Lawrence}, J.~K., {Topka}, K.~P., \& {Jones}, H.~P. 1993, \jgr, 98, 18911

\bibitem[{{Lin}(1995)}]{Lin95}
{Lin}, H. 1995, \apj, 446, 421

\bibitem[{{Lites}(2002)}]{Lites02}
{Lites}, B.~W. 2002, \apj, 573, 431

\bibitem[{{Lites} {et~al.}(2001){Lites}, {Elmore}, \& {Streander}}]{HinodeSP}
{Lites}, B.~W., {Elmore}, D.~F., \& {Streander}, K.~V. 2001, in Astronom. Soc.
  Pacific Conf. Ser., Vol. 236, Advanced Solar Polarimetry -- Theory,
  Observation, and Instrumentation, ed. {M.~Sigwarth}, 33

\bibitem[{{Lites} {et~al.}(2008){Lites}, {Kubo}, {Socas-Navarro}, {Berger},
  {Frank}, {Shine}, {Tarbell}, {Title}, {Ichimoto}, {Katsukawa}, {Tsuneta},
  {Suematsu}, {Shimizu}, \& {Nagata}}]{Lites08}
{Lites}, B.~W., {Kubo}, M., {Socas-Navarro}, H., {et~al.} 2008, \apj, 672, 1237

\bibitem[{{Lites} {et~al.}(2004){Lites}, {Scharmer}, {Berger}, \&
  {Title}}]{Lites04}
{Lites}, B.~W., {Scharmer}, G.~B., {Berger}, T.~E., \& {Title}, A.~M. 2004,
  \solphys, 221, 65

\bibitem[{{Martinez Pillet} {et~al.}(2010){Martinez Pillet}, {del Toro
  Iniesta}, {Alvarez-Herrero}, {Domingo}, {Bonet}, {Gonzalez Fernandez}, {Lopez
  Jimenez}, {Pastor}, {Gasent Blesa}, {Mellado}, {Piqueras}, {Aparicio},
  {Balaguer}, {Ballesteros}, {Belenguer}, {Bellot Rubio}, {Berkefeld},
  {Collados}, {Deutsch}, {Feller}, {Girela}, {Grauf}, {Heredero}, {Herranz},
  {Jeronimo}, {Laguna}, {Meller}, {Menendez}, {Morales}, {Orozco Suarez},
  {Ramos}, {Reina}, {Ramos}, {Rodriguez}, {Sanchez}, {Uribe-Patarroyo},
  {Barthol}, {Gandorfer}, {Knoelker}, {Schmidt}, {Solanki}, \& {Vargas
  Dominguez}}]{Imax10}
{Martinez Pillet}, V., {del Toro Iniesta}, J.~C., {Alvarez-Herrero}, A.,
  {et~al.} 2010, Sol. Phys., in press

\bibitem[{{Morinaga} {et~al.}(2008){Morinaga}, {Sakurai}, {Ichimoto},
  {Yokoyama}, {Shimojo}, \& {Katsukawa}}]{Morinaga08}
{Morinaga}, S., {Sakurai}, T., {Ichimoto}, K., {et~al.} 2008, \aap, 481, L29

\bibitem[{{Muller} \& {Roudier}(1984)}]{Muller84}
{Muller}, R. \& {Roudier}, T. 1984, \solphys, 94, 33

\bibitem[{{Narayan} \& {Scharmer}(2010)}]{Narayan10}
{Narayan}, G. \& {Scharmer}, G.~B. 2010, \aap, 524, A3+

\bibitem[{{Neckel} \& {Labs}(1994)}]{Neckel94}
{Neckel}, H. \& {Labs}, D. 1994, \solphys, 153, 91

\bibitem[{{Orozco Su{\'a}rez} {et~al.}(2007){Orozco Su{\'a}rez}, {Bellot
  Rubio}, {Del Toro Iniesta}, {Tsuneta}, {Lites}, {Ichimoto}, {Katsukawa},
  {Nagata}, {Shimizu}, {Shine}, {Suematsu}, {Tarbell}, \& {Title}}]{Orozco07}
{Orozco Su{\'a}rez}, D., {Bellot Rubio}, L.~R., {Del Toro Iniesta}, J.~C.,
  {et~al.} 2007, \pasj, 59, 837

\bibitem[{{Rabin}(1992{\natexlab{a}})}]{Rabin92b}
{Rabin}, D. 1992{\natexlab{a}}, \apjl, 390, L103

\bibitem[{{Rabin}(1992{\natexlab{b}})}]{Rabin92a}
{Rabin}, D. 1992{\natexlab{b}}, \apj, 391, 832

\bibitem[{{Rempel}(2006)}]{Rempel06}
{Rempel}, M. 2006, \apj, 647, 662

\bibitem[{{Riethm{\"u}ller} {et~al.}(2010){Riethm{\"u}ller}, {Solanki},
  {Mart{\'{\i}}nez Pillet}, {Hirzberger}, {Feller}, {Bonet}, {Bello
  Gonz{\'a}lez}, {Franz}, {Sch{\"u}ssler}, {Barthol}, {Berkefeld}, {del Toro
  Iniesta}, {Domingo}, {Gandorfer}, {Kn{\"o}lker}, \&
  {Schmidt}}]{Riethmueller10}
{Riethm{\"u}ller}, T.~L., {Solanki}, S.~K., {Mart{\'{\i}}nez Pillet}, V.,
  {et~al.} 2010, \apjl, 723, L169

\bibitem[{{R{\"o}hrbein} \& {Sch{\"u}ssler}(2011)}]{Roehrbein11}
{R{\"o}hrbein}, D. \& {Sch{\"u}ssler}, M. 2011, \aap, {in prep}

\bibitem[{{Schnerr} \& {Spruit}(2010)}]{Schnerr10}
{Schnerr}, R.~S. \& {Spruit}, H.~C. 2010, ArXiv e-prints

\bibitem[{{Schuessler} \& {Solanki}(1988)}]{Schuessler88}
{Schuessler}, M. \& {Solanki}, S.~K. 1988, \aap, 192, 338

\bibitem[{{Sch{\"u}ssler}(1992)}]{Schuessler_rev92}
{Sch{\"u}ssler}, M. 1992, in NATO Advanced Study Institute Series C Proc. 373:
  The Sun: A Laboratory for Astrophysics, ed. J.~T. {Schmelz} \& J.~C. {Brown},
  p. 191

\bibitem[{{Sheeley}(1969)}]{Sheeley69}
{Sheeley}, Jr., N.~R. 1969, \solphys, 9, 347

\bibitem[{{Solanki}(1993)}]{Solanki_rev93}
{Solanki}, S.~K. 1993, Space Science Reviews, 63, 1

\bibitem[{{Solanki} {et~al.}(2010){Solanki}, {Barthol}, {Danilovic}, {Feller},
  {Gandorfer}, {Hirzberger}, {Riethm{\"u}ller}, {Sch{\"u}ssler}, {Bonet},
  {Mart{\'{\i}}nez Pillet}, {del Toro Iniesta}, {Domingo}, {Palacios},
  {Kn{\"o}lker}, {Bello Gonz{\'a}lez}, {Berkefeld}, {Franz}, {Schmidt}, \&
  {Title}}]{SUNRISE_res}
{Solanki}, S.~K., {Barthol}, P., {Danilovic}, S., {et~al.} 2010, \apjl, 723,
  L127

\bibitem[{{Solanki} {et~al.}(2006){Solanki}, {Inhester}, \&
  {Sch{\"u}ssler}}]{Solanki_rev06}
{Solanki}, S.~K., {Inhester}, B., \& {Sch{\"u}ssler}, M. 2006, Rep. Prog.
  Phys., 69, 563

\bibitem[{{Solanki} \& {Stenflo}(1984)}]{Solanki84}
{Solanki}, S.~K. \& {Stenflo}, J.~O. 1984, \aap, 140, 185

\bibitem[{{Solanki} {et~al.}(1996){Solanki}, {Zufferey}, {Lin}, {R{\"u}edi}, \&
  {Kuhn}}]{Solanki96}
{Solanki}, S.~K., {Zufferey}, D., {Lin}, H., {R{\"u}edi}, I., \& {Kuhn}, J.~R.
  1996, \aap, 310, L33

\bibitem[{{Spruit}(1976)}]{Spruit76}
{Spruit}, H.~C. 1976, \solphys, 50, 269

\bibitem[{{Spruit}(2003)}]{Spruit03}
{Spruit}, H.~C. 2003, \solphys, 213, 1

\bibitem[{{Spruit} \& {Zwaan}(1981)}]{Spruit81}
{Spruit}, H.~C. \& {Zwaan}, C. 1981, \solphys, 70, 207

\bibitem[{{Steiner}(2005)}]{Steiner05}
{Steiner}, O. 2005, \aap, 430, 691

\bibitem[{{Steiner}(2007)}]{Steiner07}
{Steiner}, O. 2007, in Modern Solar Facilities - Advanced Solar Science, ed.
  F.~{Kneer}, K.~G. {Puschmann}, \& A.~D. {Wittmann} (Universit{\"a}tsverlag
  G{\"o}ttingen), 321

\bibitem[{{Stenflo}(1973)}]{Stenflo73}
{Stenflo}, J.~O. 1973, \solphys, 32, 41

\bibitem[{{Stenflo}(2008)}]{Stenflo_rev08}
{Stenflo}, J.~O. 2008, J. Astrophys. Astron., 29, 19

\bibitem[{{Stenflo} \& {Harvey}(1985)}]{Stenflo85}
{Stenflo}, J.~O. \& {Harvey}, J.~W. 1985, \solphys, 95, 99

\bibitem[{{Suematsu} {et~al.}(2008){Suematsu}, {Tsuneta}, {Ichimoto},
  {Shimizu}, {Otsubo}, {Katsukawa}, {Nakagiri}, {Noguchi}, {Tamura}, {Kato},
  {Hara}, {Kubo}, {Mikami}, {Saito}, {Matsushita}, {Kawaguchi}, {Nakaoji},
  {Nagae}, {Shimada}, {Takeyama}, \& {Yamamuro}}]{HinodeSOT2}
{Suematsu}, Y., {Tsuneta}, S., {Ichimoto}, K., {et~al.} 2008, \solphys, 249,
  197

\bibitem[{{Title} \& {Berger}(1996)}]{Title96}
{Title}, A.~M. \& {Berger}, T.~E. 1996, \apj, 463, 797

\bibitem[{{Title} {et~al.}(1989){Title}, {Tarbell}, {Topka}, {Ferguson},
  {Shine}, \& {SOUP Team}}]{Title89}
{Title}, A.~M., {Tarbell}, T.~D., {Topka}, K.~P., {et~al.} 1989, \apj, 336, 475

\bibitem[{{Title} {et~al.}(1992){Title}, {Topka}, {Tarbell}, {Schmidt},
  {Balke}, \& {Scharmer}}]{Title92}
{Title}, A.~M., {Topka}, K.~P., {Tarbell}, T.~D., {et~al.} 1992, \apj, 393, 782

\bibitem[{{Topka} {et~al.}(1992){Topka}, {Tarbell}, \& {Title}}]{Topka92}
{Topka}, K.~P., {Tarbell}, T.~D., \& {Title}, A.~M. 1992, \apj, 396, 351

\bibitem[{{Topka} {et~al.}(1997){Topka}, {Tarbell}, \& {Title}}]{Topka97}
{Topka}, K.~P., {Tarbell}, T.~D., \& {Title}, A.~M. 1997, \apj, 484, 479

\bibitem[{{Tsuneta} {et~al.}(2008){Tsuneta}, {Ichimoto}, {Katsukawa}, {Nagata},
  {Otsubo}, {Shimizu}, {Suematsu}, {Nakagiri}, {Noguchi}, {Tarbell}, {Title},
  {Shine}, {Rosenberg}, {Hoffmann}, {Jurcevich}, {Kushner}, {Levay}, {Lites},
  {Elmore}, {Matsushita}, {Kawaguchi}, {Saito}, {Mikami}, {Hill}, \&
  {Owens}}]{HinodeSOT}
{Tsuneta}, S., {Ichimoto}, K., {Katsukawa}, Y., {et~al.} 2008, \solphys, 249,
  167

\bibitem[{{Uitenbroek} \& {Tritschler}(2006)}]{Uitenbroek06}
{Uitenbroek}, H. \& {Tritschler}, A. 2006, \apj, 639, 525

\bibitem[{{Viticchi{\'e}} {et~al.}(2010){Viticchi{\'e}}, {Del Moro},
  {Criscuoli}, \& {Berrilli}}]{Viticchie10}
{Viticchi{\'e}}, B., {Del Moro}, D., {Criscuoli}, S., \& {Berrilli}, F. 2010,
  \apj, 723, 787

\bibitem[{{V{\"o}gler} \& {Sch{\"u}ssler}(2003)}]{Voegler03}
{V{\"o}gler}, A. \& {Sch{\"u}ssler}, M. 2003, Astronom. Nachr., 324, 399

\bibitem[{{V{\"o}gler} {et~al.}(2005){V{\"o}gler}, {Shelyag}, {Sch{\"u}ssler},
  {Cattaneo}, {Emonet}, \& {Linde}}]{Voegler05}
{V{\"o}gler}, A., {Shelyag}, S., {Sch{\"u}ssler}, M., {et~al.} 2005, \aap, 429,
  335

\bibitem[{{Walton} {et~al.}(2003){Walton}, {Preminger}, \&
  {Chapman}}]{Walton03}
{Walton}, S.~R., {Preminger}, D.~G., \& {Chapman}, G.~A. 2003, \apj, 590, 1088

\bibitem[{{Zakharov} {et~al.}(2007){Zakharov}, {Gandorfer}, {Solanki}, \&
  {L{\"o}fdahl}}]{Zakharov07}
{Zakharov}, V., {Gandorfer}, A., {Solanki}, S.~K., \& {L{\"o}fdahl}, M. 2007,
  \aap, 461, 695

\bibitem[{{Zayer} {et~al.}(1990){Zayer}, {Stenflo}, {Keller}, \&
  {Solanki}}]{Zayer90}
{Zayer}, I., {Stenflo}, J.~O., {Keller}, C.~U., \& {Solanki}, S.~K. 1990, \aap,
  239, 356

\bibitem[{{Zhao} \& {Kosovichev}(2004)}]{Zhao04}
{Zhao}, J. \& {Kosovichev}, A.~G. 2004, \apj, 603, 776

\end{thebibliography}

\end{document}